\documentclass[pdflatex,sn-mathphys]{sn-jnl}

\usepackage{aas_macros}

\jyear{2021}%

\theoremstyle{thmstyleone}%
\theoremstyle{thmstyletwo}%

\usepackage{mdframed}

\theoremstyle{thmstylethree}%
\usepackage{amsmath}
\usepackage{ragged2e}

\usepackage{csquotes}

\usepackage{xcolor}
\usepackage{multirow}
\usepackage{makecell}
\usepackage[table]{xcolor}
\usepackage{amsmath, amssymb, amsfonts}
\usepackage{bm}
\usepackage[normalem]{ulem}

\definecolor{phelipeorange}{RGB}{230,120,20}

\definecolor{berblue}{RGB}{20,80,150}

\usepackage{verbatim}

\begin{document}
\title[
Interpretable EoS from AI]{
\centering{
Physics-guided discovery of dynamical dark-energy equations of state through iterative AI reasoning}
}

\author*[1]{\fnm{Clecio R.} \sur{Bom}}\email{debom@cbpf.br}

\author*[1]{\fnm{Bernardo} \sur{M. Fraga}}\email{bernardo@cbpf.br}
\author[2,3,4]{\fnm{Miguel A.} \sur{Sabogal}}\email{miguel.sabogalgarcia@unitn.it}
\author[2]{\fnm{Armando} \sur{Bernui}}\email{bernui@on.br}

\author[1]{\fnm{Phelipe} \sur{Darc}}\email{phelipedarc@gmail.com}

\author[5,1]{\fnm{Gustavo} \sur{Schwarz}}\email{gustavo.b.schwarz@gmail.com}

\affil[1]{\orgname{Artificial Intelligence for Physics Laboratory, Centro Brasileiro de Pesquisas F\'isicas}, \orgaddress{\street{Rua Dr. Xavier Sigaud 150}, \city{Rio de Janeiro}, \postcode{22290-180}, \state{RJ}, \country{Brazil}}}
\affil[2]{ \orgname{Observatório Nacional}, \orgaddress{\street{
Rua General José Cristino 77, São Cristóvão}, \city{Rio de Janeiro}, \postcode{20921-400}, \state{RJ}, \country{Brazil}}}
\affil[3]{\orgname{Department of Physics, University of Trento}, \orgaddress{\street{Via Sommarive 14}, \postcode{38123 Povo (TN)}, \state{Trento}, \country{Italy}}}
\affil[4]{\orgname{Trento Institute for Fundamental Physics and Applications (TIFPA)-INFN}, \orgaddress{\street{Via Sommarive 14}, \postcode{38123 Povo (TN)}, \state{Trento}, \country{Italy}}}
\affil[5]{ \orgname{Instituto de Astronomia, Geofísica e Ciências Atmosféricas, Universidade de São Paulo, IAG-USP}, \orgaddress{\street{
Rua do Matão 1226, Cidade Universitária, São Paulo}, \city{São Paulo}, \postcode{05508-090}, \state{SP}, \country{Brazil}}}




\abstract{

Phenomenological model building has traditionally relied on human reasoning: equations are proposed from theoretical intuition, analogy, or empirical convenience, and only then tested against data. Here we show that this cycle can be recast as an iterative AI reasoning process for dynamical dark energy. Our framework uses a large language model to propose equations of state together with cosmological rationales, grounded by retrieval from the dark-energy literature and refined through autonomous evaluation. Each candidate is embedded in a cosmological model, optimized against observations, and assessed using likelihood performance and theoretical consistency. An independent language-model critic scores the physical motivation, novelty, clarity, stability and implementation validity of both the equation and its rationale, allowing subsequent proposals to evolve jointly in mathematical structure and
physical reasoning. Applied to cosmological data combinations
including supernovae, baryon acoustic oscillations and Planck likelihoods, the
framework identifies two parameterizations that, to the best of our knowledge,
have not previously been explored and that are competitive with established
forms. For Pantheon+ supernovae, DESI DR2 baryon acoustic oscillations and the full Planck 2018 temperature, polarization, and lensing likelihoods, the best AI-selected
model attains larger Bayesian evidence than the traditional
parameterizations considered here by more than one unit.
These results show that AI-guided reasoning can complement physical model building by proposing and evaluating interpretable phenomenological parameterizations for dynamical dark energy.

}

\keywords{Artificial intelligence, Cosmology, Dark Energy, Observational Cosmology, Bayesian model comparison, Equation-of-state parameterizations}



\maketitle

\section{Introduction}

\label{sec1}

\begin{figure}[h!]
\centering
\includegraphics[width=1\linewidth]{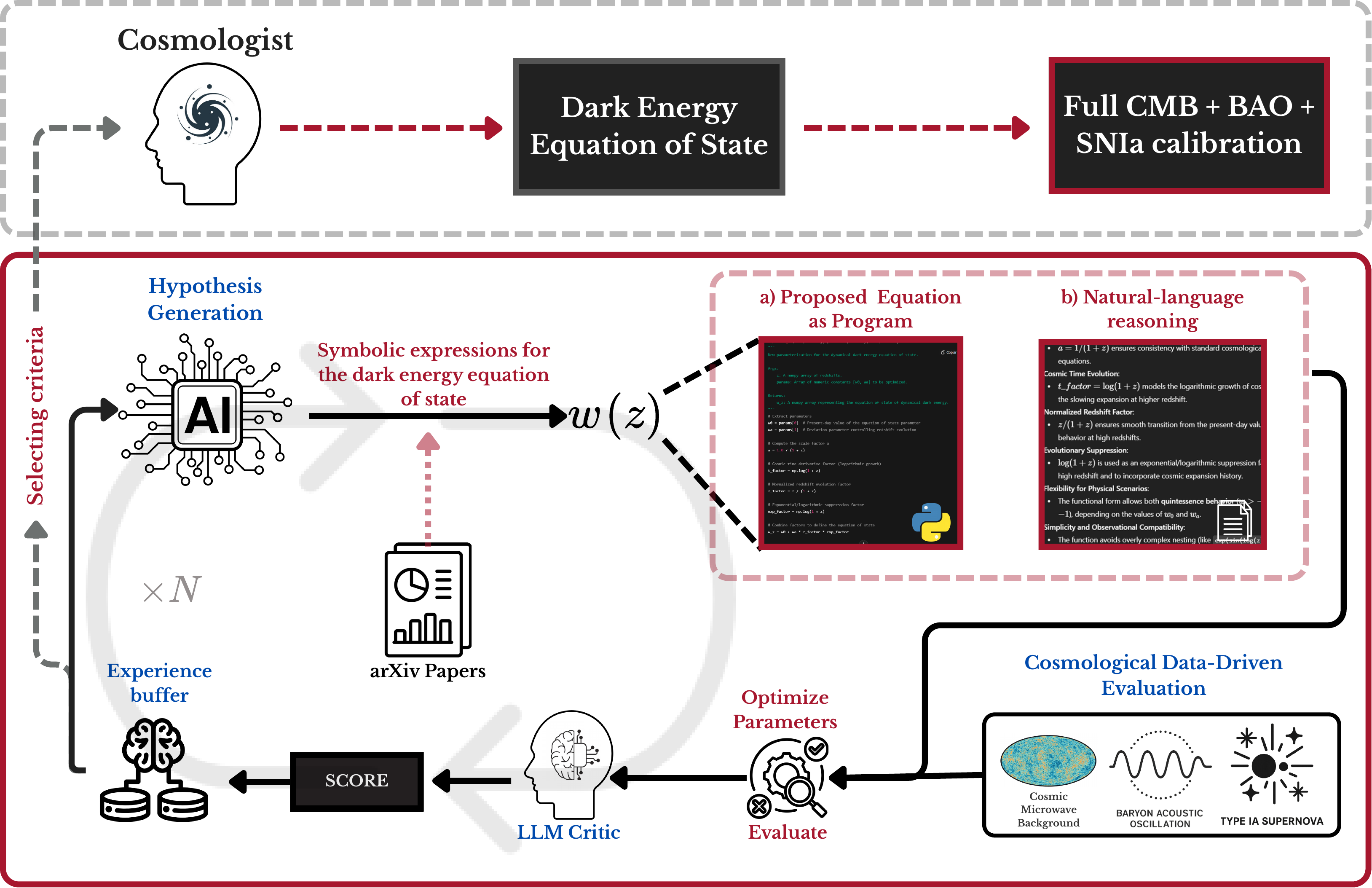}
\caption{\textbf{Workflow for AI-reasoned discovery of dark-energy equations of state.}
A cosmologist defines the target datasets and physical selection criteria for viable dynamical dark-energy models. The hypothesis generator LLM model proposes symbolic parameterizations of \(w(z)\) as executable Python functions together with natural-language cosmological rationales. Candidate generation is conditioned on retrieved arXiv literature, cosmological context, and high-performing examples stored in an experience buffer. Each proposal is embedded in the cosmological model, optimized against supernova, baryon acoustic oscillation, and compressed cosmic microwave background data, and assigned a statistical score. An independent critic LLM model evaluates the physical motivation, novelty, clarity, stability, and implementation validity of the equation and its rationale, and scores it. The combined score updates the experience buffer, selecting hypotheses for subsequent iterations. The loop therefore evolves both functional forms and physics-based reasoning, so that discovered equations are judged by their propagated impact on cosmological observables rather than by direct fitting of \(w(z)\) alone.}

\label{fig:cosmoAI}
\end{figure}

The discovery that the expansion of the Universe is accelerating ~\citep{riess1998observational,perlmutter1999measurements} provided the first compelling evidence, within the framework of general relativity, for a dominant component with negative pressure, commonly referred to as dark energy.
Within the standard cosmological interpretation, this component 
contributes approximately $70\%$ of the present cosmic energy budget. 
The concordance $\Lambda$CDM model describes dark energy as a cosmological constant, $\Lambda$, corresponding to a constant equation of state (EoS) parameter $w\equiv p/\rho=-1$.

Despite its empirical success, $\Lambda$CDM leaves unresolved profound theoretical questions concerning the physical nature of dark energy, such as why its energy density has the observed magnitude, or why it becomes dynamically dominant only at recent cosmic times~\citep{weinberg1989cosmological,peebles2003cosmological}.
These issues have motivated extensive exploration of dynamical dark-energy models, in which the EoS departs from $w=-1$ and evolves as a function of cosmic time, scale factor, or redshift. 
Examples include scalar-field models such as quintessence, non-canonical kinetic constructions such as k-essence, interacting or coupled dark-energy scenarios, early dark-energy models, and phenomenological parameterizations designed to capture deviations from $\Lambda$CDM~\citep{ratra1988cosmological,caldwell1998,z1999PhRvL..82..896Z,2000PhRvL..85.4438A,linder2003exploring}.

Recent results from the Dark Energy Spectroscopic Instrument (DESI) have substantially sharpened observational tests of late-time cosmic acceleration. Its first-year Baryon Acoustic Oscillation (BAO) analysis and the subsequent DR2 constraints, based on the first three years of observations, provide the most precise BAO constraints to date across a broad redshift range extending to the Ly-$\alpha$ forest at $z\sim 4$~\citep{desi2024bao,DESI2025}. When combined with Type Ia supernova (SNIa) compilations~\citep{brout2022pantheonplus,des2024sn} and cosmic microwave background (CMB) constraints~\citep{planck2020overview}, these measurements remain broadly consistent with flat-$\Lambda$CDM model, but they also exhibit model-dependent preferences for extensions in which the dark energy EoS evolves with redshift. 
In particular, DESI DR2 reports that combinations of BAO with CMB and supernova data can favor $w_0w_a$CDM-like descriptions over $\Lambda$CDM at a significance that depends on the supernova sample and analysis assumptions~\citep{DESI2025,desi2025extendedde,Oliveira2025a,Oliveira2025b,Giare2024,Sabogal2025,Silva2025,Colgain2026}.

Nevertheless, if substantiated by future DESI releases and by independent probes from forthcoming surveys, such a deviation would have profound implications for the physical interpretation of cosmic acceleration. It would suggest that dark energy is not fully captured by a cosmological constant, but instead may reflect a dynamical component, an effective description of new gravitational physics, or a more complex sector whose phenomenology is only partially represented by current low-dimensional parameterizations.
These developments highlight a central limitation of parameterized dark-energy inference: the evidence for evolution is necessarily conditioned on the functional form assumed for $w(z)$. 
Current analyses typically rely on a small set of low-dimensional, hand-crafted parameterizations, especially CPL-like models, which offer interpretability 
and statistical control but explore only a restricted region of the possible EoS landscape. 
If the underlying dark-energy phenomenology departs from these templates, existing searches may underestimate the evidence for dynamics, force the signal into an inadequate functional form, or obscure the physical mechanism responsible for the observed expansion history.

A natural way forward is, therefore, to enlarge the space of EoS candidates while retaining the interpretability and statistical control that make phenomenological models useful. Flexible reconstructions, including binned, Gaussian-process, and other non-parametric approaches, reduce assumptions about the shape of \(w(z)\), but they do not generally yield compact analytic forms that can be directly interpreted, compared with theoretical mechanisms, or propagated into future analyses. 
Conversely, traditional phenomenological parameterizations are analytically transparent, but their functional choices remain human-designed and sample only a restricted subset of possible dynamical behaviors. 
This situation motivates a discovery framework that can search over symbolic forms of \(w(z)\), while using physical reasoning, observational performance, theoretical consistency, and implementation validity as selection criteria.

Large language models (LLMs) and related agentic AI systems offer a promising route towards such a framework. Recent advances in AI-assisted science, including symbolic equation discovery with inductive biases~\citep{cranmer2020discovering,ruan2026discovering}, theory-guided scientific discovery that combines data with background knowledge~\citep{2023NatCo..14.1777C}, LLM-guided equation search~\citep{llmsr}, and emerging agentic workflows for scientific ideation, experimentation and evaluation~\citep{wang2023scientific,lu2026towards,shao2026sciscigpt,zhang2025exploring}, suggest that these systems can contribute to early-stage model building by synthesizing scientific context, proposing candidate structures and articulating their motivations. Yet this promise comes with a verification problem: AI-generated hypotheses may be compelling in language while remaining weakly grounded empirically or theoretically, and can place additional burden on downstream validation and review~\citep{lu2026towards,zhang2025exploring}. Harnessing their generative capacity for cosmology, therefore, requires a controlled loop in which hypotheses are motivated, implemented, tested, criticized and refined.

A second concern is more specific to the use of language models in theoretical
or phenomenological physics. Because such models are trained on large
scientific corpora, apparent discovery may reduce to interpolation among
well-known templates rather than exploration of genuinely distinct functional
structures. For dark energy this is a sharp issue: the literature is dominated
by \(\Lambda\)CDM, \(w\)CDM, Chevalier-Polarski-Linder (CPL) and a small number of related two-parameter
ansätze. A useful AI-assisted model-building procedure must therefore be tested
not only on real data, but also on controlled cases where the target expansion
history is intentionally outside this familiar template class.

Here we introduce a physics-guided AI reasoning framework for discovering dynamical dark-energy EoS. The method combines LLM generation of symbolic candidate functions, retrieval-augmented access to the dark-energy literature, numerical optimization within a cosmological inference pipeline, and an independent LLM Critic that evaluates physical plausibility, novelty, clarity of reasoning, stability, and implementation validity. Each candidate is treated as a complete cosmological hypothesis: a functional form for \(w(z)\), a physics-based rationale, optimized parameters, and quantitative performance against data. Successful hypotheses are stored in an experience buffer and used to condition subsequent generations, guiding later generations toward stronger equations and rationales.

\newpage
\section{Results}
\label{sec:results}

\subsection{AI-reasoned workflow for discovering dark-energy parameterizations}
\label{sec:workflow}

We first define the discovery problem at the level of cosmological observables rather than at the level of a reconstructed EoS. 
As summarized in Figure~\ref{fig:cosmoAI}, a cosmologist specifies the target datasets. The hypothesis-generator model then proposes symbolic forms for the dark-energy EoS, \(w(z)\), as executable Python functions, together with a natural-language rationale grounded in retrieved dark-energy literature and in high-performing candidates stored in an experience buffer.

Each proposed equation is embedded in the background cosmological model and its free parameters are optimized against the selected SNIa, BAO, and CMB datasets. 

Candidate models are therefore scored by their propagated effect on $\chi^2$ over the cosmological datasets, distances and CMB summary quantities, not by direct regression to a target \(w(z)\). 
An independent language-model critic evaluates the results over the data, the physical motivation, novelty, clarity, numerical stability, and implementation validity of each proposal, after instructions on the physical criteria needed. The critic also provides suggestions for modifying the current proposal. The statistical and reasoning-based scores update the experience buffer, allowing subsequent iterations to evolve both the mathematical structure of \(w(z)\) and its cosmological justification.

This iterative loop procedure is designed to avoid hard-coding the standard cosmological model as the preferred outcome. To test this point, and before analyzing real data, we perform a blind validation on simulated observations generated from an intentionally non-standard dark-energy cosmology.

\subsection{Blind validation on an exotic mock cosmology}
\label{sec:exotic-mock}

We generated mock supernova, two-dimensional BAO and compressed CMB observables from an exotic flat universe cosmology with 
\(\Omega_{\rm m}=0.6\) and
\(H_0=90~{\rm km\,s^{-1}\,Mpc^{-1}}\). The mock dark-energy equation of state was chosen to differ strongly from standard late-time parameterizations, that is 
\begin{equation}
w_{\rm exotic}(z)=\frac{\alpha}{1+z^\beta} \,,
\label{eq:mock-param}
\end{equation}
with \(\alpha=-5\) and \(\beta=2\). This choice provides a compact stress test: it is analytically simple, yet sufficiently far from CPL-like evolution to reveal whether the pipeline can adapt to non-standard expansion histories. Gaussian noise was added with a fractional dispersion of approximately \(6\%\) for the BAO observables and a fixed magnitude uncertainty 
\(\sigma_{\rm SN}=0.18~{\rm mag}\) for the supernova sample. 
The simulated supernova catalog span 
\(0.01 \leq z \leq 2.5\), with \(N=1000\) logarithmically spaced redshifts.

The blind test evaluates whether the discovery procedure can identify viable dynamical dark-energy behavior when the data are generated far from the Planck-preferred flat-\(\Lambda\)CDM region. Candidate models were evaluated using $\chi^2$ goodness of fit to the mock observables, numerical stability, and consistency with physically motivated viability conditions.

This mock was designed as a test of ansatz dependence, not as a realistic
dark-energy theory. Its expansion history is far from \(\Lambda\)CDM and
CPL-like evolution, so recovery cannot be attributed to reuse of familiar
templates. The test therefore probes whether the search can shift its
functional preferences when the observables demand non-standard dynamics.

The results are summarized in Table~\ref{tab:combined_results}. We retained three representative AI-generated candidates spanning distinct transition structures. 
None is an exact symbolic recovery of the injected functional form given in Equation~(\ref{eq:mock-param}); nevertheless, the best candidates reproduce the expansion history, with a strongly evolving late-time equation of state that weakens toward higher redshift encoded in the mock observables. 
This provides a controlled check that the search is not restricted to standard dark-energy forms. The full validation analysis is presented in Appendix~\ref{appendix-B}.

As observed in Figure~\ref{fig:mock_cosmo} in Appendix, 
the CPL ansatz becomes structurally inadequate. 
It absorbs the non-standard evolution of \(w(z)\) through biased background parameters, whereas the best AI-generated models recover the injected \(\Omega_{\rm m}\)--\(H_0\) region within the quoted credible intervals. In particular, AI-sim2 gives
\(\Omega_{\rm m}=0.61^{+0.13}_{-0.13}\) and
\(H_0=88.6^{+11.1}_{-8.0}\,{\rm km\,s^{-1}\,Mpc^{-1}}\).

For the exotic-mock test, we compute Bayes factors relative to the injected functional form, \(\ln\mathcal{B}_{Ei}\equiv \ln Z_E-\ln Z_i\), where \(Z_E\) is the evidence of the model using the true exotic equation of state and \(Z_i\) is the evidence of candidate \(i\). Thus, positive values favour the true
exotic model. On the Kass--Raftery scale, CPL and AI-sim3 are strongly
disfavoured (\(\ln\mathcal{B}_{Ei}>6\)), while AI-sim1 and AI-sim2 show only
moderate evidence tension with the injected model
(\(\ln\mathcal{B}_{Ei}<3\)).
Nevertheless, already at iteration 100, AI-sim3 recovers the underlying background cosmology with good accuracy despite not being favored in Bayesian evidence relative to CPL at this stage. This early recovery is initially accompanied by broader posterior distributions than those obtained with CPL; however, later iterations progressively reduce these uncertainties toward values comparable to those obtained when fitting the simulated data with the true EoS. In contrast, CPL yields substantially tighter constraints but remains strongly biased with respect to the injected cosmology, failing to recover the true background parameters within high credibility regions. Subsequent iterations (226 for AI-sim1 and 410 for AI-sim2) refine both the recovered cosmology and the statistical support, becoming strongly preferred over CPL while remaining only moderately disfavored relative to the true model.

The predictive metrics show the same behaviour at the observable level: the AI-generated models reduce the supernova and BAO residuals relative to CPL, and AI-sim2 matches the injected-form fit (termed ExoFit in Table~\ref{tab:combined_results}) for the BAO observables within the precision of the mock analysis. 

The best AI-generated equations follow the effective evolution of \(w_{\rm exotic}(z)\) across the calibrated redshift range, yielding Hubble diagram and BAO predictions that are nearly indistinguishable from those of the injected model at the resolution shown. The validation therefore demonstrates recovery of the observable cosmological content, rather than exact recovery of the generating formula.

The selectivity of this mock recovery is also relevant. In the exotic-mock experiment, more than
\(10^3\) candidate equations were generated and evaluated by the same discovery
loop, but only two achieved \(\ln\mathcal{B}_{Ei}<3\) relative to the injected
functional form. As a null control, we applied the same pre-specified procedure
to mock data generated from a CPL \(w_0w_a\)CDM cosmology, using the same
mock-data prescription as in the exotic test. The fiducial model was based on
the DESI-DR2 BAO+CMB+Pantheon+ best-fit solution,
\(H_0=67\,{\rm km\,s^{-1}\,Mpc^{-1}}\), \(\Omega_{m0}=0.32\),
\(w_0=-0.838\) and \(w_a=-0.62\)~\citep{DESI2025}. In \(10^3\) attempts, no
AI-generated candidate matched the CPL \(\chi^2\); the best candidate had a
\(\chi^2\) more than an order of magnitude larger. Together, these controls
show that the procedure adapts to non-standard expansion histories without
spuriously promoting novel forms when the data are generated by a standard
two-parameter template.

\begin{table}[ht]
\centering
\footnotesize
\caption{\textbf{
Recovery
of the Exotic Mock Cosmology:} Posterior parameter estimates for the Exotic Mock Cosmology (ExoMCMC) fitted in the investigated data using the CPL 
parametrization, as well as those proposed by the 
AI framework, and the difference in log Bayesian evidence to the fitted injected form. 
The injected simulated values are presented in the first column (ExoC). 
Each model was fitted to the simulated SNIa, BAO, and CMB datasets. 
Quoted uncertainties correspond to the $68\%$ credible intervals. 
The units of $H_0$ are $\mathrm{km\,s^{-1}\,Mpc^{-1}}$.}
\label{tab:combined_results}
\setlength{\tabcolsep}{3pt}
\begin{tabular}{lcccccc}
\hline \hline
& ExoC & ExoFit & CPL & AI-sim1(226) & AI-sim2(410) 
& AI-sim3(100)\\
\hline \hline
\multicolumn{7}{c}{\textit{Posterior Parameters}} \\
\hline

$\Omega_M$ 
& $0.60$ 
& $0.52^{+0.10}_{-0.10}$ 
& $0.81^{+0.05}_{-0.05}$ 
& $0.58^{+0.13}_{-0.06}$ 
& $0.61^{+0.13}_{-0.13}$ 
& $0.55^{+0.18}_{-0.16}$\\

$H_0$ 
& $90.0$ 
& $96.3^{+10.9}_{-8.2}$ 
& $77.9^{+2.6}_{-2.1}$ 
& $91.2^{+5.0}_{-8.5}$ 
& $88.6^{+11.1}_{-8.0}$ 
& $93.8^{+18.0}_{-12.5}$\\


$\ln \mathcal{B}_{E i}$
& $-$
& $0.0$
& $6.97$
& $2.24$
& $1.55$
& $6.34$ \\

\hline
\multicolumn{7}{c}{\textit{Performance of the \textrm{EoS} proposed on fitting the exotic mock data}} \\
\hline
MSE$_\mathrm{SN}$ 
& $-$ & 0.037 & 0.053 & 0.034 & 0.033 & 0.036\\

MAE$_\mathrm{SN}$ 
& $-$ & 0.155 & 0.185 & 0.148 & 0.147 & 0.152\\

$R^2_\mathrm{SN}$ 
& $-$ & 0.9966 & 0.9952 & 0.9969 & 0.9970 & 0.9967\\

MSE$_\mathrm{BAO}$ 
& $-$ & 0.532 & 0.807 & 0.558 & 0.526 & 0.566\\

MAE$_\mathrm{BAO}$ 
& $-$ & 0.487 & 0.644 & 0.492 & 0.486 & 0.516\\

$R^2_\mathrm{BAO}$ 
& $-$ & 0.9880 & 0.9818 & 0.9874 & 0.9882 & 0.9872\\

\hline
\end{tabular}
\end{table}



\vspace{0.3cm}

\newpage

\newpage

\subsection{Equation-of-state discovery in real data}
\label{sec:real_data}

Through successive reasoning--evaluation cycles, the pipeline generated approximately $3\times10^3$ candidate equations of state, with each iteration informed by the statistical performance and critic feedback of earlier proposals. The discovery process was guided by a composite objective combining cosmological fit, physical motivation, originality, numerical robustness, and implementation validity. The critic model operates within this loop procedure, evaluating each proposal and its rationale in conjunction with its $\chi^2$ performance across the current cosmological datasets: weakly model-dependent transverse BAO measurements (hereafter \texttt{2D BAO}), \texttt{Pantheon+} supernovae, and compressed CMB constraints through the shift parameter $R$ and acoustic scale $\ell_A$. We refer to this reduced combination as the discovery dataset.

The highest-ranked 25 candidates were inspected by domain experts
before full inference. No retained model showed implementation artefacts or
unstable phenomenology. For every retained candidate, the minimum \(\chi^2\)
on \texttt{Pantheon+}+\texttt{2D BAO}+\(\{R,\ell_A\}\) was lower than that obtained with
CPL or JBP on the same dataset. 
This discovery dataset was used only for screening; final performance was
assessed with Bayesian evidence from the full-likelihood validation analyses
below.

The final analysis separates discovery from validation. The 25 candidates
selected with the discovery dataset were re-evaluated with the
full Planck 2018 likelihood and with DESI-DR2, Union3 and DESY5 combinations.
Only two remained within approximately one log-evidence unit of CPL for
\texttt{DESY5+DESI-DR2+Planck 2018}. These two parameterizations, denoted AI~1 and
AI~2, were then carried forward to the full model comparison in
Table~\ref{tab:combined_results_performance}; the selection sequence is
summarized in Table~\ref{tab:selection_criteria}. Evidence in
Table~\ref{tab:combined_results_performance} is reported as
\(\ln\mathcal{B}_{\Lambda i}=\ln Z_{\Lambda{\rm CDM}}-\ln Z_i\), with negative
values favouring model \(i\) over \(\Lambda\)CDM. Comparisons among dynamical
forms follow from differences in these log evidences and are conditional on
the selected fixed parameterizations, not trials-corrected significances for
the AI search procedure. We now describe these two forms and their
phenomenological behaviour.

\begin{table}[ht]

\caption{\textbf{Discovery and validation of AI-generated parameterizations.}
Candidate ranking used the reduced discovery likelihood and LLM-critic scores;
Bayesian evidence was reserved for the final full-likelihood validation.}
\label{tab:selection_criteria}
\centering
\small
\setlength{\tabcolsep}{5pt}
\begin{tabular}{p{0.25\linewidth}p{0.53\linewidth}c}
\hline
Step & Criterion & Candidates \\
\hline
AI generation
&  Executable \(w(z)\) proposals accompanied by physical reasoning;
&  $3\times10^3$ \\

AI-critic ranking
& Composite score from \(\chi^2\) on \texttt{Pantheon+}+\texttt{2D BAO}+\(\{R,\ell_A\}\),
novelty, physical motivation, numerical stability and implementation validity. Each retained model had a lower minimum \(\chi^2\) than CPL or JBP on the
discovery data;
&  $25$ \\

Expert inspection
& Human review for implementation artefacts or unstable phenomenology;

& $25$ \\

Full-likelihood validation
& Bayesian evidence within \(\sim 1\) log-evidence unit of CPL, or better,
on \texttt{Planck 2018+\({\rm DESI}\)-DR2+DESY5}.
& 2 \\
\hline
\end{tabular}
\end{table}

\begin{table*}[t]
\centering
\footnotesize
\caption{\justifying
Post-selection model-comparison results for the standard dark-energy parameterizations and the two AI-generated parameterizations, shown for each dataset combination. We report the difference in minimum chi-squared relative to $\Lambda$CDM, $\Delta\chi^{2}_{\rm min}\equiv \chi^{2}_{{\rm min},i}-\chi^{2}_{{\rm min},\Lambda{\rm CDM}}$, the logarithm of the Bayes factor, $\ln \mathcal{B}_{\Lambda i}$, and the corresponding conservative evidence-based significance in units of $\sigma$. Negative values of $\Delta\chi^{2}_{\rm min}$ indicate an improved best fit relative to $\Lambda$CDM, whereas negative values of $\ln \mathcal{B}_{\Lambda i}$ indicate Bayesian preference for the dynamical dark-energy model. The arrows mark the direction of preference: $(\uparrow)$ denotes preference for the dynamical dark-energy model and $(\downarrow)$ for $\Lambda$CDM. Shaded entries highlight the strongest-performing model within each class for a given dataset combination.}

\renewcommand{\arraystretch}{1.5}
\setlength{\tabcolsep}{0.2pt}
\label{tab:combined_results_performance}
\resizebox{\textwidth}{!}{%
\begin{tabular}{l||
*{4}{>{\centering\arraybackslash}p{1.9cm}}||
*{2}{>{\centering\arraybackslash}p{1.9cm}}}
\hline
 & \multicolumn{4}{c||}{\textbf{Traditional Parametrizations}} 
 & \multicolumn{2}{c}{\textbf{Novel AI Parametrizations}} \\
\hline

\textbf{Parameter} & \textbf{CPL} \cite{chevallier2001accelerating, linder2003exploring} & \textbf{JBP}\cite{JBP} & \textbf{EXP}\cite{dimakis2016general}& \textbf{BA}\cite{BA} & \textbf{AI 1} & \textbf{AI 2} \\
\hline \hline

\hline
&\multicolumn{6}{c}{\textit{weakly-model-dependent 2D BAO $+$ Planck 2018}} \\
\hline

$\Delta\chi^{2}_{\rm min}$ 
& $-2.72$ & \cellcolor{gray!20} $-4.40$ & $-2.86$ & $-3.64$ 
& \cellcolor{gray!20} $-3.20$ & $-2.40$ \\
$\ln \mathcal{B}_{\Lambda i}$ 
& $-4.24$ & $-4.59$ & $-3.83$ & \cellcolor{gray!20} $-6.75$ 
& $-3.94$ & \cellcolor{gray!20} $-4.39$ \\
$\mathrm{Significance}\,[\sigma]$ 
& $3.38\,(\uparrow)$ & $3.49\,(\uparrow)$ & $3.24\,(\uparrow)$ & \cellcolor{gray!20} $4.09\,(\uparrow)$ 
& $3.28\,(\uparrow)$ & \cellcolor{gray!20} $3.43\,(\uparrow)$ \\

\hline
&\multicolumn{6}{c}{\textit{weakly-model-dependent 2D BAO $+$ PantheonPlus SNIa $+$ Planck 2018}} \\
\hline

$\Delta\chi^{2}_{\rm min}$ 
& $-5.86$ & $-4.76$ & $-7.34$ & \cellcolor{gray!20} $-7.66$ 
& \cellcolor{gray!20} $-7.02$ & $-6.86$\\
$ \ln \mathcal{B}_{\Lambda i}$ 
& $-3.67$ & \cellcolor{gray!20} $-3.89$ & $-3.71$ & $-2.93$ 
& \cellcolor{gray!20} $-4.88$ & $-3.62$ \\
$\mathrm{Significance}\,[\sigma]$ 
& $3.19\,(\uparrow)$ & \cellcolor{gray!20} $3.26\,(\uparrow)$ & $3.20\,(\uparrow)$ & $2.92\,(\uparrow)$ 
& \cellcolor{gray!20} $3.58\,(\uparrow)$ & $3.17\,(\uparrow)$  \\

\hline
&\multicolumn{6}{c}{\textit{weakly-model-dependent 2D BAO $+$ Union3 SNIa $+$ Planck 2018}} \\
\hline

$\Delta\chi^{2}_{\rm min}$ 
& $-11.34$ & $-11.36$ & \cellcolor{gray!20} $-12.56$ & $-12.36$ 
& \cellcolor{gray!20} $-11.68$ & $-11.40$ \\
$\ln \mathcal{B}_{\Lambda i}$ 
& \cellcolor{gray!20} $-6.45$ & $-6.12$ & $-6.02$ & $-5.59$ 
& $-6.05$ & \cellcolor{gray!20} $-6.20$ \\
$\mathrm{Significance}\,[\sigma]$ 
& \cellcolor{gray!20} $4.01\,(\uparrow)$ & $3.92\,(\uparrow)$ & $3.90\,(\uparrow)$ & $3.78\,(\uparrow)$ 
& $3.91\,(\uparrow)$ & \cellcolor{gray!20} $3.95\,(\uparrow)$ \\

\hline 
&\multicolumn{6}{c}{\textit{weakly-model-dependent 2D BAO $+$ DESY5 SNIa $+$ Planck 2018}} \\
\hline

$\Delta\chi^{2}_{\rm min}$ 
& $-22.60$ & \cellcolor{gray!20} $-24.04$ & $-21.96$ & $-21.58$ 
& $-21.30$ & \cellcolor{gray!20} $-23.94$\\
$\ln \mathcal{B}_{\Lambda i}$ 
& $-11.64$ & \cellcolor{gray!20} $-12.19$ & $-11.13$ & $-10.69$ 
& $-10.29$ & \cellcolor{gray!20} $-12.06$\\
$\mathrm{Significance}\,[\sigma]$ 
& $5.19\,(\uparrow)$ & \cellcolor{gray!20} $5.30\,(\uparrow)$ & $5.09\,(\uparrow)$ & $4.99\,(\uparrow)$ 
& $4.91\,(\uparrow)$ & \cellcolor{gray!20} $5.27\,(\uparrow)$\\

\hline\hline\hline
&\multicolumn{6}{c}{\textit{DESI-DR2 BAO $+$ Planck 2018}} \\
\hline

$\Delta\chi^{2}_{\rm min}$ 
& $-9.06$ & $-4.46$ & $-9.04$ & \cellcolor{gray!20} $-9.70$ 
& $-8.02$ & \cellcolor{gray!20} $-10.64$\\
$\ln \mathcal{B}_{\Lambda i}$ 
& \cellcolor{gray!20} $0.24$ & $2.41$ & $0.49$ & $0.92$ 
& $2.72$ & \cellcolor{gray!20} $0.68$\\
$\mathrm{Significance}\,[\sigma]$ 
& \cellcolor{gray!20} $1.42\,(\downarrow)$ & $2.72\,(\downarrow)$ & $1.66\,(\downarrow)$ & $1.97\,(\downarrow)$ 
& $2.81\,(\downarrow)$ & \cellcolor{gray!20} $1.81\,(\downarrow)$\\

\hline
&\multicolumn{6}{c}{\textit{DESI-DR2 BAO $+$ PantheonPlus SNIa $+$ Planck 2018}} \\
\hline

$\Delta\chi^{2}_{\rm min}$ 
& $-11.30$ & $-6.38$ & \cellcolor{gray!20} $-11.36$ & $-9.92$ 
& \cellcolor{gray!20}$-11.48$ & $-8.58$\\
$\ln \mathcal{B}_{\Lambda i}$ 
& \cellcolor{gray!20} $1.28$ & $2.24$ & $1.38$ & $2.34$ 
& \cellcolor{gray!20}$-1.17$ & $2.13$\\
$\mathrm{Significance}\,[\sigma]$ 
& \cellcolor{gray!20} $2.19\,(\downarrow)$ & $2.64\,(\downarrow)$ & $2.24\,(\downarrow)$ & $2.69\,(\downarrow)$ 
& \cellcolor{gray!20}$2.13\,(\uparrow)$ & $2.60\,(\downarrow)$\\

\hline
&\multicolumn{6}{c}{\textit{DESI-DR2 BAO $+$ Union3 SNIa $+$ Planck 2018}} \\
\hline

$\Delta\chi^{2}_{\rm min}$ 
& $-15.10$ & $-12.38$ & $-15.04$ & \cellcolor{gray!20} $-15.22$ 
& \cellcolor{gray!20} $-14.88$ & $-13.22$\\
$\ln \mathcal{B}_{\Lambda i}$ 
& \cellcolor{gray!20} $-2.29$ & $-1.09$ & $-1.49$ & $-1.26$ 
& \cellcolor{gray!20} $-3.79$ & $-1.89$\\
$\mathrm{Significance}\,[\sigma]$ 
& \cellcolor{gray!20} $2.67\,(\uparrow)$ & $2.08\,(\uparrow)$ & $2.30\,(\uparrow)$ & $2.17\,(\uparrow)$ 
& \cellcolor{gray!20} $3.23\,(\uparrow)$ & $2.49\,(\uparrow)$ \\ 

\hline
& \multicolumn{6}{c}{\textit{DESI DR2 BAO + DESY5 SNIa $+$ Planck 2018}} \\
\hline

$\Delta\chi^{2}_{\rm min}$ 
& $-28.82$ & $-27.96$ & \cellcolor{gray!20} $-30.70$ & $-28.12$ 
& \cellcolor{gray!20} $-28.04$ & $-27.96$\\
$\ln \mathcal{B}_{\Lambda i}$ 
& \cellcolor{gray!20} $-9.61$ & $-9.00$ & $-8.80$ & $-8.30$ 
& \cellcolor{gray!20} $-9.44$ & $-9.19$\\
$\mathrm{Significance}\,[\sigma]$ 
& \cellcolor{gray!20} $4.77\,(\uparrow)$ & $4.63\,(\uparrow)$ & $4.59\,(\uparrow)$ & $4.47\,(\uparrow)$ 
& \cellcolor{gray!20} $4.73\,(\uparrow)$ & $4.67\,(\uparrow)$ \\

\hline

\hline

\hline
\end{tabular}
}
\end{table*}

\vspace{0.5cm}
\noindent \textbf{AI 1: A bounded quadratic dark-energy parameterization with phantom-crossing phenomenology.}
The first AI-selected model is
\begin{equation}
w(z) = -1 - w_0\left(\frac{z^2}{1+z^2}\right) + w_a\left(\frac{z^2}{1+z^2}\right)^2 - w_0 w_a \, ,
\label{eq:newEOS}
\end{equation}
which defines a bounded deformation of \(\Lambda\)CDM in terms of the compact variable
\begin{equation}
x(z)\equiv \frac{z^2}{1+z^2}\in[0,1) \, .
\end{equation}
Here \(w_0\) and \(w_a\) are phenomenological shape parameters and should not be identified with the present-day value and derivative of the equation of state in the CPL sense. Equation~(\ref{eq:newEOS}) belongs to the broader class of regular redshift-dependent dark-energy parameterizations, while extending the standard two-parameter phenomenology beyond linear forms such as CPL and beyond rational ans\"atze such as JBP and Barboza--Alcaniz. Expressed in \(x\), the model is a quadratic deformation built on a smooth bounded transition kernel related to those used in steep equation-of-state parameterizations \citep{jaber2018probing}. Its distinctive feature is the specific nonlinear structure: a linear term in \(x\), a quadratic term in \(x^2\), and the constant offset \(-w_0w_a\). Together these define a bounded two-parameter ansatz with finite asymptotic limits and sufficient flexibility to accommodate qualitatively different phantom-crossing trajectories. To our knowledge, this exact form has not previously been studied as a dark-energy equation-of-state parameterization.

The limiting values are
\begin{equation}
w(0) = -1 - w_0 w_a \, ,
\end{equation}
and
\begin{equation}
w(z)\rightarrow -1 - w_0 + w_a - w_0 w_a \qquad (z\rightarrow\infty) \, ,
\end{equation}
so the model interpolates smoothly between two finite asymptotic regimes. This controlled asymptotic behaviour is one of its main advantages: unlike parameterizations that continue to drift at large redshift, Eq.~(\ref{eq:newEOS}) remains bounded while still permitting a crossing of the phantom divide within the observationally relevant regime.

\begin{figure}
   \centering
   \includegraphics[width=0.49\linewidth]{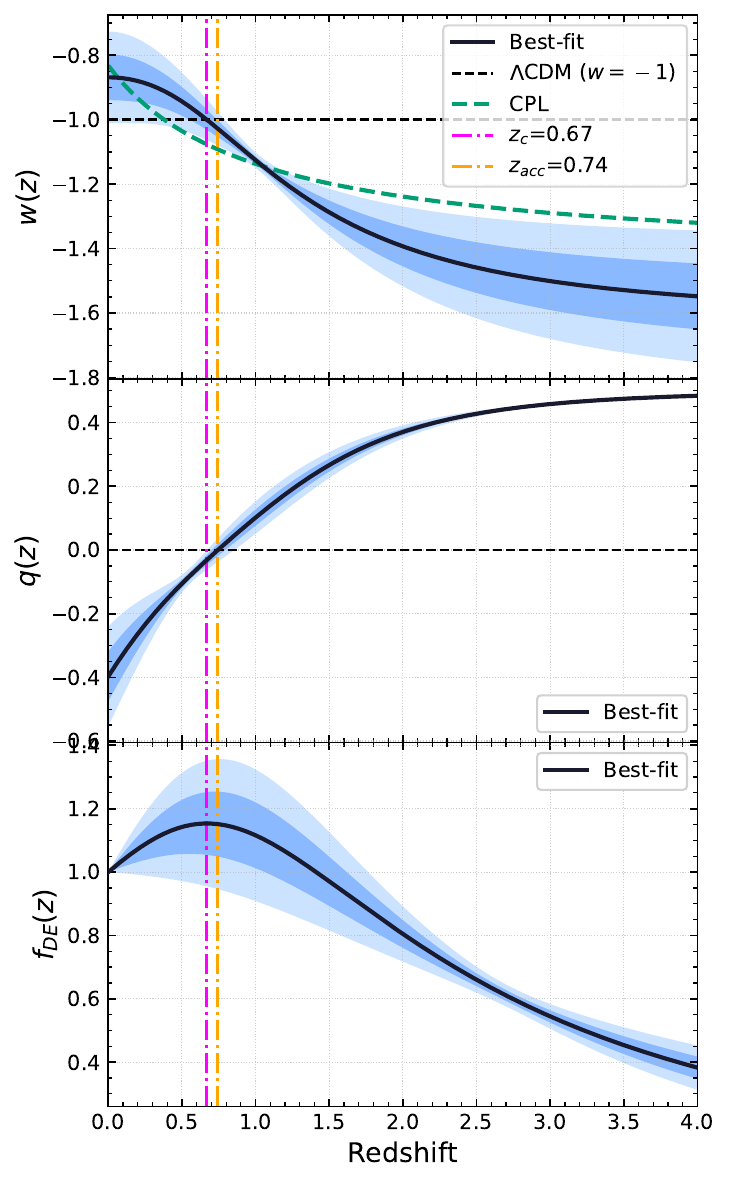}
    \includegraphics[width=0.49\linewidth]{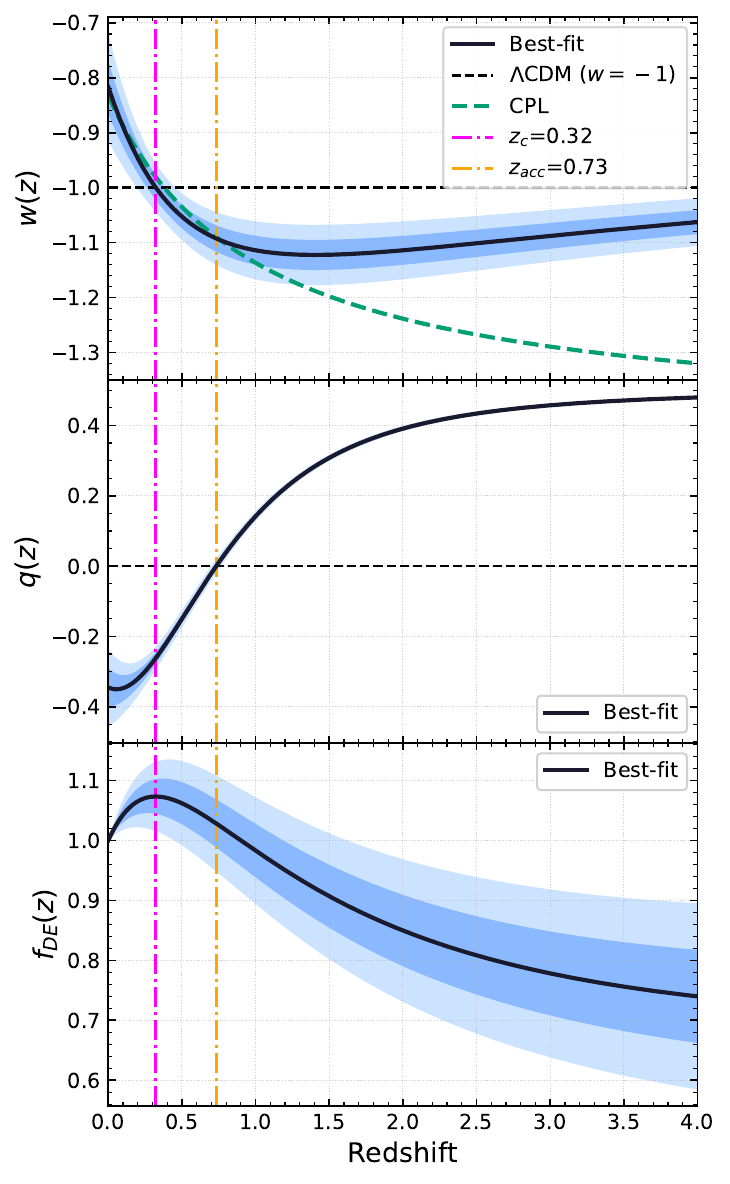}
   \caption{Equation of state, energy density and deceleration parameter for the AI 1 (Left) and AI 2 (Right) model. The solid black line indicates the best-fit values while the blue bands represent the $68\%$ and $95\%$ confidence intervals. The orange vertical dash-dotted lines indicates the redshift of phantom crossing ($z_c$), and the onset of acceleration ($z_{acc}$). The green dashed line is the CPL parametrization. The values were computed using the parameters obtained from \texttt{Planck}, \texttt{2D BAO} and \texttt{Pantheon Plus}.}
    \label{fig:eos_panel}
\end{figure}

AI~1 admits both phantom-to-non-phantom and non-phantom-to-phantom trajectories. The best-fit reconstruction shown in Fig.~\ref{fig:eos_panel} follows the former, with \(w(z)<-1\) at earlier times and \(w(z)>-1\) at lower redshift. This should not be interpreted as an early phase of dark-energy domination. Even when the equation of state is phantom-like, the dark-energy density remains subdominant to matter and radiation over the corresponding redshift range, so the background expansion does not enter premature acceleration.

\vspace{0.5cm}
\noindent \textbf{AI 2: A damped intermediate-redshift dark-energy parameterization with controlled asymptotics.}
The second AI-selected model is
\begin{equation}
w(z)=w_0+w_a\,\frac{z}{1+z}\,
\exp\!\left[-\left(\frac{z}{1+z}\right)^2(1+z)^{1/4}\right] \, ,
\label{eq:new_eos}
\end{equation}
which may be interpreted as an exponentially damped deformation of the CPL form. 
Although AI~2 is algebraically close to CPL, it imposes a different
phenomenological prior on the expansion history. CPL ties the low-redshift
slope to the high-redshift limit \(w_0+w_a\), whereas AI~2 damps this
deformation and restores \(w(z\to\infty)=w_0\). The model therefore tests
whether the data prefer transient late-time freedom without an accompanying
early-time asymptotic shift. The redshift dependence retains the \(z/(1+z)\)
kernel, but the dynamical correction is exponentially localized. The resulting
equation of state is regular for \(z\geq0\) and bounded at both low and high
redshift.

For \(z\ll1\), Eq.~(\ref{eq:new_eos}) reduces to
\begin{equation}
w(z)\simeq w_0+w_a z + \mathcal{O}(z^2) \, ,
\end{equation}
thereby preserving the leading low-redshift behaviour constrained by supernova and BAO distance measurements. In the opposite limit,
\begin{equation}
w(z)\rightarrow w_0 \qquad (z\rightarrow\infty) \, ,
\end{equation}
because the exponential factor removes the dynamical correction at early times. Unlike CPL, whose high-redshift limit is \(w_0+w_a\), AI~2 therefore retains late-time flexibility without imposing a fixed asymptotic shift. In this parametrization, \(w_a\) controls the amplitude of a transient deformation, whereas \(w_0\) fixes both the present-day and early-time equation of state.

A notable feature of Eq.~(\ref{eq:new_eos}) is that the correction term vanishes in both asymptotic regimes: at \(z=0\) through the prefactor \(z/(1+z)\), and at large redshift through exponential damping. The model thus describes a bounded departure from a constant-\(w\) background, confined to an intermediate-redshift window. To our knowledge, this exact form has not previously been studied as a dark-energy equation-of-state parametrization, although it is naturally connected to the broader class of bounded and transition-like phenomenological ans\"atze.

The best-fit reconstruction in Fig.~\ref{fig:eos_panel} illustrates this structure directly. AI~2 departs from $w$CDM over a limited low-to-intermediate-redshift interval, while remaining consistent with a decaying dark-energy fraction towards the past. Its interest lies in combining observationally accessible late-time dynamics with a controlled high-redshift limit.

\begin{figure}[htpb!]
    \centering
    \includegraphics[width=0.9\textwidth]{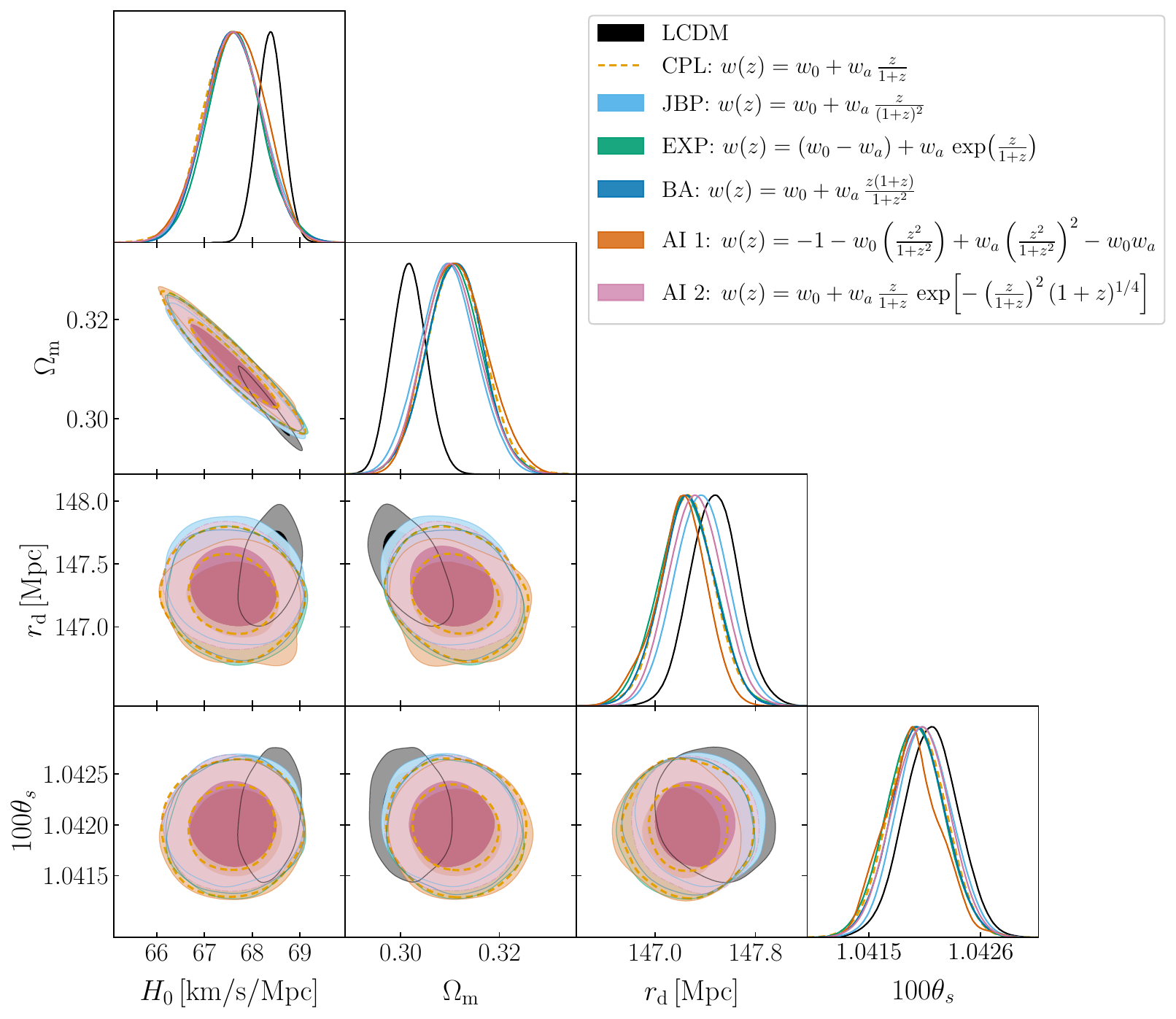}
    \caption{ \justifying Marginalized one-dimensional posterior distributions and contours (68\% and 95\% CL) for the parameters, $H_{0}$, $\Omega_{\rm m}$, $w_0$, and $w_a$ from CMB measurements joint with DESI-DR2 data and its combination with PantheonPlus. }
    \label{fig:trinagleplot}
\end{figure}
\begin{figure}[htpb!]
    \centering
    \includegraphics[width=0.45\textwidth]{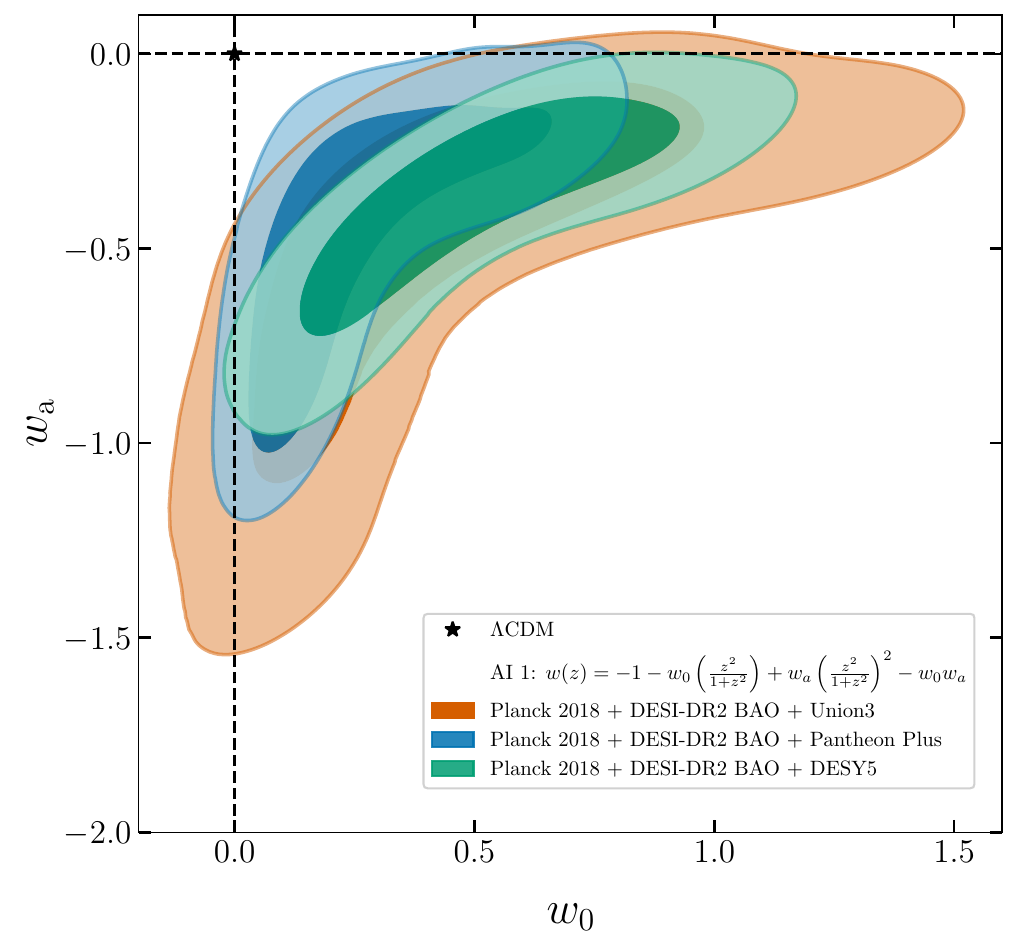}
    \includegraphics[width=0.45\textwidth]{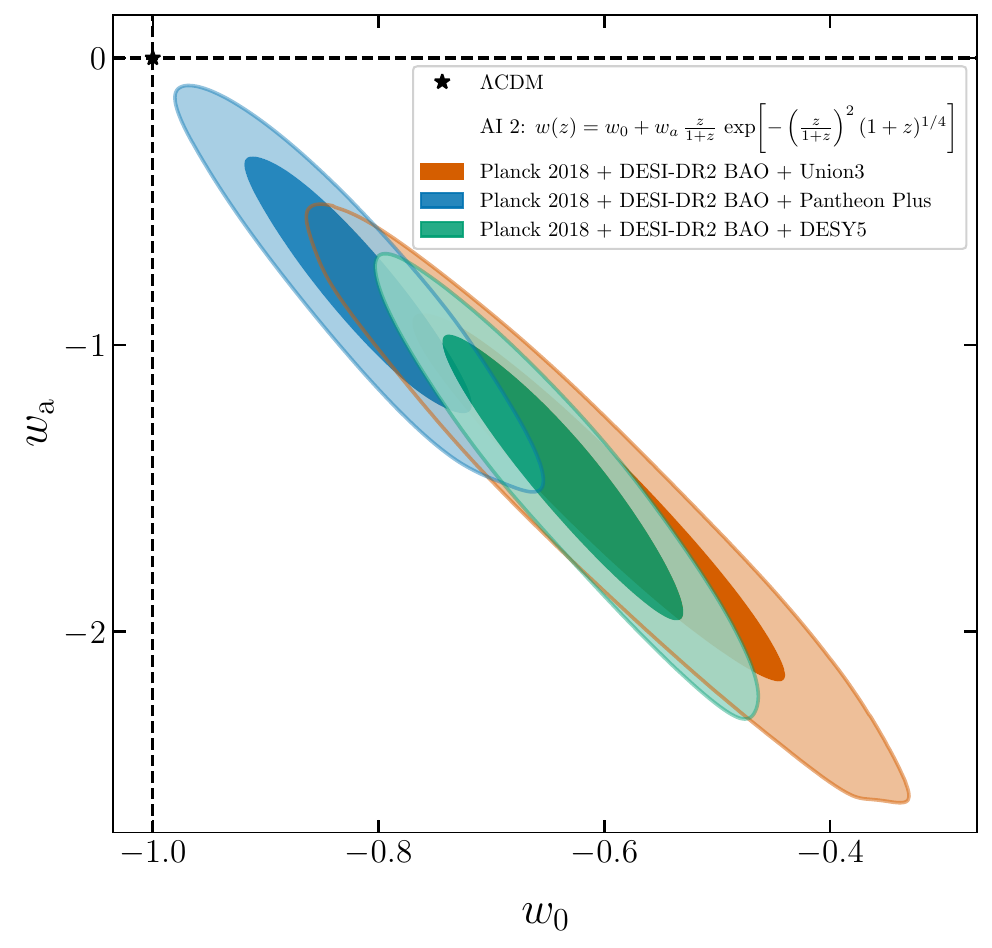}
    \caption{\justifying Marginalized one-dimensional posterior distributions and two-dimensional contours (68\% and 95\% CL) for the parameters $H_{0}$, $\Omega_{\rm m}$, $w_0$, and $w_a$, in the context of the AI 1 (left panel) and AI 2 (right panel) parametrizations. Results are shown for combinations of CMB and DESI-DR2 with each of the three SNIa samples.}
    \label{fig:w0wa}
\end{figure}

\vspace{0.5cm}
\noindent \textbf{Observational behaviour of the AI-discovered models.}
\linebreak
Although AI 1 and AI 2 realize different forms of late-time dynamics, they leave the early-Universe sector largely unchanged relative to the standard phenomenological families. In the DESI-DR2 BAO + Planck 2018 + PantheonPlus combination, both models preserve the Planck-constrained values of $10^{-2}\omega_b$, $\omega_{\rm cdm}$, $100\theta_s$, and $r_d$ at the level of the traditional parameterizations, while the main differences appear in the dark-energy sector and in correlated late-time parameters (Table~\ref{table:combined_colored} in Appendix~\ref{appendix-A}). The same trend is visible in the joint posterior projections, where the AI models remain embedded in the same $(H_0,\Omega_{\rm m},r_d,100\theta_s)$ manifold as CPL, JBP, EXP, and BA, rather than opening a qualitatively distinct region of background parameter space 
(Figure~\ref{fig:trinagleplot}). 
At the same time, the two AI forms reshape the dark-energy sector differently: AI 2 retains a CPL-like anti-correlation in the $(w_0,w_a)$ plane, whereas AI~1 compresses the preferred region into a narrower band, reflecting its non-linear mapping between parameters and expansion history (Figure~\ref{fig:w0wa}).

Figure~\ref{fig:eos_panel} shows the redshift evolution of the equation of state, $w(z)$, the deceleration parameter, $q(z)$, and the normalized dark-energy density, $f_{\rm DE}(z)$, for AI 1 and AI 2. The top, middle and bottom rows show $w(z)$, $q(z)$ and $f_{\rm DE}(z)$, respectively, with AI 1 shown in the left column and AI 2 in the right column. In both models, the EoS is quintessence-like today and exhibits a phantom phase in the past. For AI~2, this phase is bounded, with the EoS returning to quintessence-like behaviour at $z_c \simeq 8$, depending on the dataset combination. For AI~1, the phantom regime persists to the early Universe. Despite this phantom behaviour, the dark-energy density remains sufficiently subdominant that accelerated expansion is not induced during the early-time phantom phase.

Despite their different EoS evolution, the two models yield similar acceleration histories, with a transition from decelerated to accelerated expansion at $z_{\rm acc} \simeq 0.7$. This shows that $q(z)$ is relatively insensitive to the detailed functional form of $w(z)$ within the range allowed by the data. The confidence bands for $q(z)$ are also narrower than those for $w(z)$, indicating that the observations constrain the expansion history more tightly than the detailed redshift dependence of the EoS.

Both AI 1 and AI 2 are robust across dataset combinations. Considering all eight combinations, the onset of acceleration lies in the range $z_{\rm acc}\in[0.72,0.80]$ for AI 2 and $z_{\rm acc}\in[0.72,0.84]$ for AI 1. By contrast, the low-redshift phantom-crossing redshift shows larger scatter, with $z_c\in[0.26,0.52]$ for AI 2 and $z_c\in[0.50,0.75]$ for AI 1. This again indicates that the data constrain the background expansion more tightly than the detailed EoS shape.

\par
\begin{figure}[htpb!]
    \centering
    \includegraphics[width=0.95\textwidth]{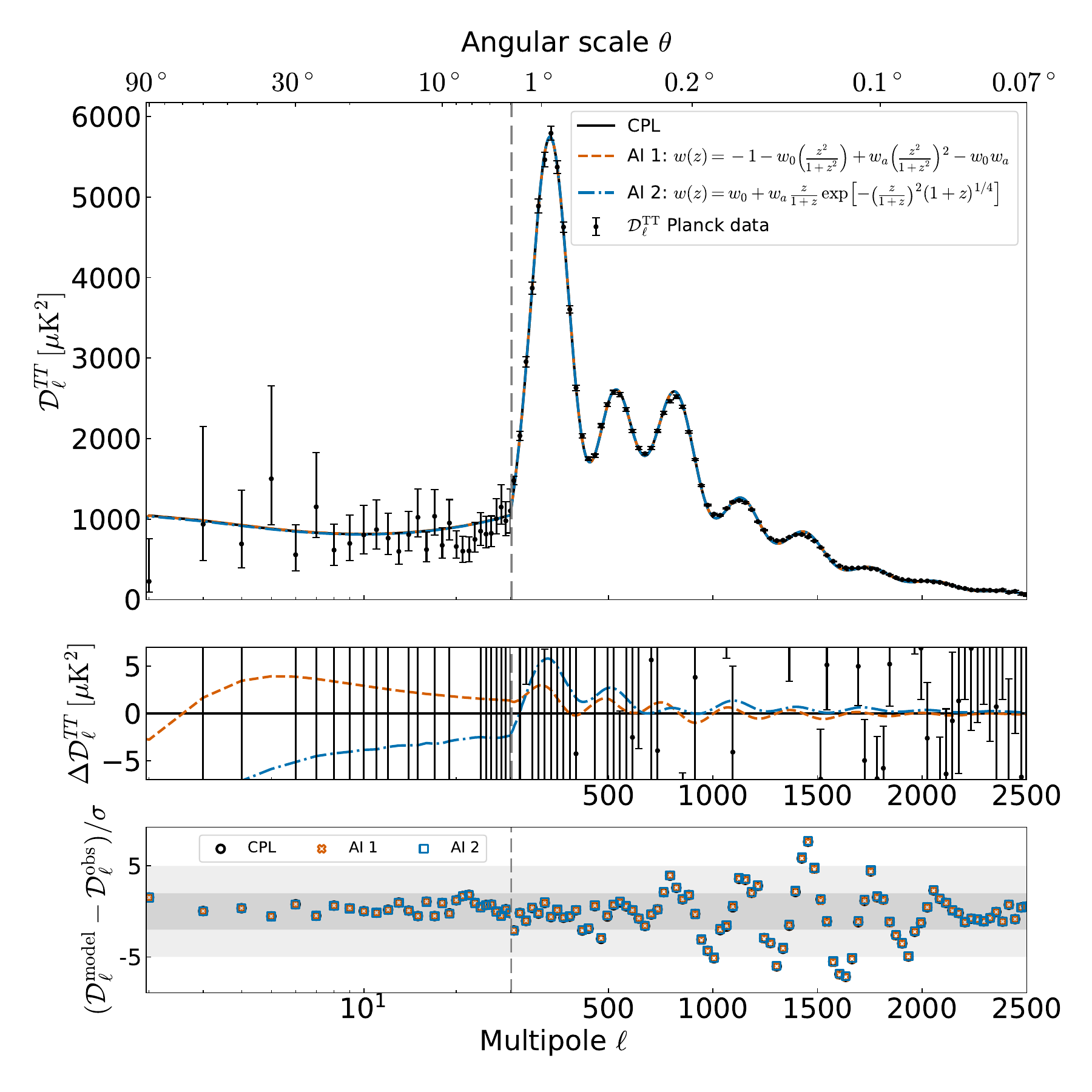}
    \caption{\justifying Comparison between the CPL and AI parametrizations (AI1 and AI2) in terms of their theoretical predictions for the CMB temperature anisotropy power spectrum, with all cosmological parameters fixed to their respective best-fit values obtained from the joint CMB + DESI-DR2 + PantheonPlus analysis. The error bars on the Planck data points correspond to $\pm 1\sigma$ uncertainties. The middle panel shows the difference between the CPL and AI predictions, while the lower panel displays the relative deviation normalized by the observational uncertainties.}
    \label{fig:CTT}
\end{figure}

 Figure~\ref{fig:CTT} shows that, at their joint \texttt{Pantheon+}+\texttt{DESI-DR2}+\texttt{Planck 2018} best-fit points, CPL and the two AI-generated parametrizations yield nearly degenerate CMB temperature spectra. The acoustic peaks are reproduced with essentially the same fidelity, and the normalized residuals show no systematic preference for any model. The remaining CPL--AI differences are small and arise primarily at low multipoles, where late-time dark-energy dynamics have their largest projected imprint on the temperature spectrum.

\newpage

\newpage


\section{Discussion}\label{sec:discussion}

The present results suggest a change in how phenomenological dark-energy models can be constructed. In the conventional workflow, an equation of state is chosen \emph{a priori}---typically for reasons of tractability, interpretability, or historical use---and only then confronted with cosmological data, as in CPL and related low-dimensional ans\"atze \cite{chevallier2001accelerating,linder2003exploring,JBP,BA}. That strategy has been highly productive, but it also fixes asymptotic behaviour and the placement of dynamical freedom in redshift before inference begins. The framework developed here reverses that logic: candidate equations are proposed together with physical arguments, evaluated through propagated observables, criticized for statistical and theoretical deficiencies, and revised iteratively. The object of exploration is therefore not a single ansatz, but the phenomenological space itself.

This places the framework outside the usual scope of classical symbolic-regression and equation-discovery methods, which are designed to recover compact expressions reproducing observed relations or learned surrogates \cite{2009Sci...324...81S,2023NatCo..14.1777C}. The present search is directed at a different target: not a function to be compressed, but a class of phenomenological hypotheses whose viability is decided only after cosmological propagation, likelihood evaluation, and critic-based assessment of physical motivation, numerical robustness, and implementation validity. In dark-energy studies, this distinction is substantial because the central difficulty is not simply to reconstruct \(w(z)\), but to determine where functional freedom can be introduced without hindering the high-redshift sector. The method is therefore better understood as physics-guided phenomenological reasoning than as direct equation fitting.

This interpretation also places the framework within the emerging literature on AI-assisted scientific discovery, where the emphasis has shifted from prediction alone to the cycle of hypothesis generation, evaluation, and revision under domain-specific constraints \cite{Krenn:2022,wang2023scientific,zhang2025exploring}. The claim here is more precise: the loop procedure does not assert autonomous scientific understanding, it makes phenomenological reasoning explicit, auditable, and revisable. Each candidate is associated with a physical rationale, executable form, critic-based assessment, and cosmological likelihood, so the route from proposal to selection can be inspected directly. The methodological advance is therefore the conversion of ansatz construction into a traceable search process, rather than automation in the abstract. Framed in this way, the contribution lies in rendering phenomenological reasoning explicit and auditable, rather than treating AI-generated outputs as evidence of scientific understanding \cite{messeri2024artificial}.

The blind mock-universe analysis controls for template dependence. The
injected cosmology was deliberately far from \(\Lambda\)CDM and CPL-like
evolution: CPL was strongly disfavoured
(\(\ln\mathcal{B}_{Ei}=6.97\) relative to the injected form), whereas the best
AI-generated candidate reached \(\ln \mathcal{B}_{Ei}=1.55\) without exactly recovering
the injected equation. Only two of more than \(10^3\) mock candidates achieved
\(\ln\mathcal{B}_{Ei}<3\). Thus the search was not confined to familiar ansätze, and
high-evidence recovery was selective rather than automatic.

The real-data results point to the same general pattern in two different ways. AI~1 and AI~2 are not equivalent parameterizations, but both place the additional dark-energy freedom in the late-time Universe while keeping the high-redshift sector controlled. AI~2 is best understood in these terms: its importance is not the large algebraic distance from CPL, but the fact that it preserves the useful low-redshift structure of CPL while removing the forced asymptotic shift to \(w_0+w_a\). It therefore decouples late-time flexibility from early-time asymptotics. AI~1 realizes a different possibility, allowing a bounded late-time phantom excursion together with a controlled asymptotic regime at earlier times. The shared feature matters more than the algebraic difference between the two forms: current background data appear to reward dark-energy freedom that is concentrated near the onset of acceleration, rather than freedom that propagates monotonically into the deep past.

The statistical results support a selective rather than universal advantage for the AI-discovered forms. AI~1 is favored over \(\Lambda\)CDM by the Bayesian evidence criterion for DESI-DR2 BAO + PantheonPlus + Planck 2018, and it provides the strongest evidence among the models considered for DESI-DR2 BAO + Union3 + Planck 2018. In other dataset combinations, the AI forms remain competitive without being uniquely preferred. What emerges is a repeated preference for bounded late-time deformations whose complexity is penalized less severely by the evidence than standard two-parameter ans\"atze when the data constrain how dark-energy freedom is organized.

These gains are not accompanied by large excursions in the parameters that anchor the early Universe solution. The stability of \(10^{-2}\omega_b\), \(\omega_{\rm cdm}\), \(100\theta_s\), \(n_s\), and \(r_d\), together with the near-degeneracy of the best-fit CMB temperature spectra across the acoustic peaks, indicates that the AI-discovered models act primarily through late-time background evolution rather than through substantial modification of recombination-era physics. Their statistical preference is therefore more naturally interpreted as a refinement of late-time phenomenology than as evidence for a qualitatively new early-Universe sector.

These equations should be interpreted phenomenologically rather than as microphysical theories of dark energy. Neither AI~1 nor AI~2 is derived from a unique scalar-field Lagrangian, effective field theory, or modified-gravity construction, and AI~1 retains the familiar caveats associated with phantom behaviour and phantom-divide crossing \cite{hu2005crossing,vikman2005can,copeland2006dynamics}. Their significance lies in showing that literature-grounded proposal, critic-based reasoning, and cosmological inference can be integrated into a reproducible search for effective models.

A final implication concerns the inductive structure of phenomenological inference. The framework does not eliminate prior assumptions about what constitutes a viable dark-energy model, it makes them more explicit. In conventional parametrization studies, much of the inductive content is concentrated in the ansatz itself, where asymptotic behaviour, transition structure, and allowable curvature are fixed before inference begins. Here those preferences are distributed across identifiable elements of the search loop---retrieval, critic criteria, executable-validity constraints, and observational scoring. The gain is not neutrality, but inspectability: one can ask which physical and statistical pressures are selecting a given class of equations, and revise those pressures as the scientific objective changes. The next task is to know whether this more explicit control remains effective once perturbation stability, growth observables, and effective-field-theory consistency conditions are incorporated directly into the search objective. 

Taken together, these tests position the framework as an auditable route to
interpretable phenomenological model construction, rather than as an
autonomous derivation of dark-energy microphysics. The exotic mock demonstrates
that the search can move beyond familiar \(\Lambda\)CDM- and CPL-like templates
when the expansion history demands it. In real data, two compact
two-parameter equations of state survive full Planck, BAO and supernova
validation and remain competitive with established parameterizations after
Bayesian complexity penalization.

\section{Methods}\label{sec:methods}

We describe in detail the procedures followed to develop the framework used for EoS discovery. 
We also present the cosmological datasets used in our analyses.


\subsection{Pipeline and experiment designs}\label{sec:pipeline}

Recent work has shown that large language models can contribute to symbolic equation discovery by generating executable mathematical forms and refining them through data-driven selection \citep{llmsr}. We situate our method within this broader class of language-model-assisted discovery, but target a different object: reasoning-guided phenomenological model construction. In our framework, a candidate is not defined only by a symbolic expression for the dark-energy equation of state, but by the expression together with an explicit physical rationale. The rationale is not used as a post hoc explanation; it is stored and evaluated later by human cosmologists to assess the level of the critique. 

The hypothesis generator, implemented with Qwen3-30B-A3B, proposes candidate equations of state, $w(z)$, as executable Python functions. For each candidate, the free parameters are optimized against the datasets described in Sections~\ref{sec:datasets} and \ref{sec:eval}. The statistical component of the score is based on the reduced $\chi^2$ normalized by the corresponding JBP value, providing a common benchmark against a standard phenomenological parameterization. The optimized parameters, statistical score, critic assessment and rationale are stored in an experience buffer for each iteration separately.

At each iteration, the generator prompt is conditioned on entries drawn from this buffer. By default, the five highest-ranked candidates are included, providing a compact memory of successful functional structures and their associated rationales. To limit premature convergence, we also allow the prompt to include randomly selected lower-ranked candidates from the history of previous attempts, thereby increasing functional diversity in the search context.
This buffer-based conditioning biases the search toward empirically and physically promising regions of hypothesis space without imposing a fixed parametric family.

The search also uses retrieval-augmented generation, described in Section~\ref{sec:RAG}. Retrieved passages from the dark-energy literature are inserted into the generation prompt and provide context for both the proposed functional form and its accompanying rationale. This step anchors the reasoning process in existing phenomenology while allowing the model to propose new parameterizations.

An independent critic, implemented with GPT-5.1 mini, evaluates each candidate after the cosmological fit has been performed. The critic assigns scores for physical plausibility, novelty, clarity of reasoning, numerical stability and implementation correctness, and provides targeted suggestions for modifying the functional form. These suggestions are incorporated into subsequent generator prompts, allowing the critic to influence both candidate ranking and the direction of later proposals. The final candidate score is computed by multiplying the statistical and critic components, and determines the ordering of the experience buffer.

Before the final discovery runs, the critic and scoring scheme were calibrated using a separate set of randomly sampled candidate evaluations that was not used in the final model comparison. These samples were reviewed by human cosmologists to assess whether the critic's judgments of physical motivation and novelty were supported by the dark-energy literature, and whether penalties were applied consistently to predefined failure modes, including weak physical motivation, lack of novelty, numerical instability and implementation errors.

The calibration was restricted to the critic prompt and to the fixed relative weights assigned to the score components. It did not use the final candidate set, did not alter the cosmological likelihoods or optimized parameters, and did not change the datasets used for model comparison. The calibrated scoring scheme was then frozen before the final discovery runs.

The resulting loop is designed to avoid selecting candidates by likelihood alone. A high-ranked candidate must improve or remain competitive with the benchmark fit, remain numerically stable, have a coherent physical rationale and differ non-trivially from established parameterizations. The procedure therefore couples statistical evaluation with structured physical critique, allowing the search to revise both the functional form and the reasoning used to motivate it.


\subsection{Retrieval-augmented generation system}
\label{sec:RAG}

To ground candidate generation in the dark-energy literature, we constructed a retrieval-augmented generation system based on arXiv manuscripts. Using the arXiv API, we assembled a final corpus of approximately 7,000 \LaTeX\ manuscripts from the cosmology and gravitation categories, filtered by metadata to retain dark-energy-related topics including \emph{dark energy models}, \emph{dynamical dark energy}, \emph{interacting dark energy}, \emph{dark-energy cosmology}, \emph{quintessence}, \emph{phantom dark energy}, \emph{dark-energy equation of state}, \emph{modified-gravity dark energy}, \emph{dark energy}, and \emph{dark-energy observations}. As an auxiliary screening step, we applied a Qwen3-32B classifier \citep{qwen3technicalreport} to the abstracts to verify that the manuscripts were consistent with these categories.

The retained manuscripts were preprocessed to remove references, author metadata, markup, and other non-content elements. The remaining text was segmented into overlapping chunks of 2,000 tokens with a 300-token overlap, preserving local scientific context while maintaining retrieval granularity. Each chunk was embedded with the \texttt{bge-1.5} model \citep{bge_embedding} into a 1,024-dimensional vector space and stored in a vector database.

\begin{figure}
    \centering
    \includegraphics[width=1\linewidth]{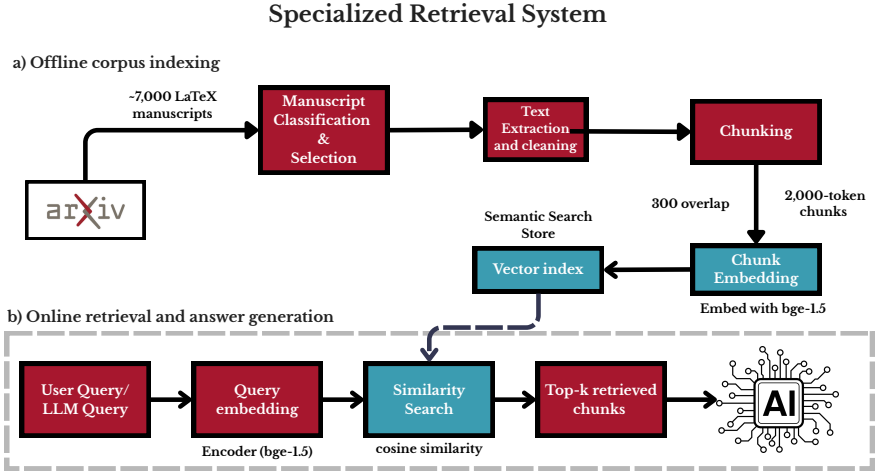}
    \caption{Construction and retrieval workflow of the pipeline domain corpus. Schematic overview showing separation between offline indexing and online inference-time retrieval. }
    \label{fig:ragpipeline}
\end{figure}

During candidate generation, the retrieval query was embedded with the same model and compared against the stored chunks using cosine similarity. The six most similar chunks were then inserted into the hypothesis-generation prompt. Retrieval therefore provided literature context for both the proposed functional form and its physical rationale, without constraining the model to reproduce existing parameterizations. Based on qualitative inspection by the authors, generations produced with retrieval appeared better grounded in the dark-energy literature and more consistent in their physical motivation than generations from the base model alone. We therefore adopted this design to improve the interpretability and domain relevance of the generated rationales while preserving open-ended exploration of the dark-energy equation-of-state space (Fig.~\ref{fig:ragpipeline}).

\subsection{Cosmological data and observables}
\label{sec:datasets}

We constrain the dark-energy models identified by our framework using complementary late- and early-universe probes: Type Ia supernova (SNIa) distance measurements, baryon acoustic oscillation (BAO) data, and Planck 2018 cosmic microwave background (CMB) temperature, polarization, and lensing likelihoods.

\paragraph{Type Ia supernovae.}
We consider three contemporary SNIa compilations. \texttt{Pantheon+} provides 1701 light curves from 1550 distinct SNe Ia over the redshift range \(0.001<z<2.26\)~\cite{Brout:2022vxf}. \texttt{Union3} contains 2087 cosmologically useful SNe Ia analysed within the UNITY1.5 Bayesian framework, which models selection effects, outliers, and population-level systematics in a unified way~\cite{Rubin:2023jdq}. \texttt{DESY5} comprises 1635 photometrically classified DES SNe Ia in the range \(0.10<z<1.13\), together with an external low-redshift sample of 194 SNe Ia spanning \(0.025<z<0.10\)~\cite{DES:2024jxu}. In the remainder of the paper, these samples are denoted \texttt{Pantheon+}, \texttt{Union3}, and \texttt{DESY5}, respectively.

\paragraph{Planck 2018 CMB.}
For the CMB we use the Planck 2018 legacy likelihoods~\cite{Planck:2018vyg,Planck:2019nip}. Our baseline combination includes the high-\(\ell\) \texttt{Plik} TT ($30 \leq \ell \leq 2508$), TE and EE ($30 \leq \ell \leq 1996$) likelihood (\texttt{Planck\_2018\_highl\_TTTEEE}), together with the low-\(\ell\) temperature and polarization likelihoods \texttt{Planck\_2018\_lowl\_TT} and \texttt{Planck\_2018\_lowl\_EE} ($2 \leq \ell \leq 29$). We also include the Planck 2018 CMB lensing likelihood, \texttt{Planck\_2018\_lensing\_PR3}, reconstructed from the temperature and polarization four-point function~\cite{Planck:2018lbu}. We refer to this combination collectively as \texttt{Planck 2018}.

\paragraph{Baryon acoustic oscillations.}
We use two BAO datasets. The first is \textbf{DESI DR2}, for which we adopt the isotropic and anisotropic BAO measurements from galaxies, quasars and Ly\(\alpha\) tracers over \(0.295\leq z\leq2.33\), expressed in terms of the transverse comoving distance $D_{\mathrm{M}}/r_{\mathrm{d}}$, the Hubble horizon $D_{\mathrm{H}}/r_{\mathrm{d}}$, and the angle-averaged distance $D_{\mathrm{V}}/r_{\mathrm{d}}$, all normalized to the comoving sound horizon at the drag epoch, $r_{\mathrm{d}}$, together with their reported covariance structure~\cite{DESI:2025zgx,DESI:2025zpo}. The second is a weakly model-dependent compilation of transverse angular BAO measurements, \(\theta_{\rm BAO}(z)\), constructed from 17 measurements spanning low to high redshift, including the correlated subsample compiled in Ref.~\cite{Menote22} and three additional independent determinations at \(z=0.11\), \(1.725\), and \(2.225\)~\cite{deCarvalho21,Avila25,deCarvalho18,Sabogal:2025qhz}. We denote these datasets by \texttt{DESI-DR2} and \texttt{2D BAO}, respectively.

\subsubsection{Background expansion and distance measures}

For any candidate equation-of-state parameterization \(w(z)\), the dark-energy contribution to the expansion history is
\begin{equation}
f_{\mathrm{DE}}(z)\equiv \frac{\rho_{\mathrm{DE}}(z)}{\rho_{\mathrm{DE},0}}
=\exp\!\left[3\int_0^z \frac{1+w(z')}{1+z'}\,dz'\right].
\end{equation}
Assuming a spatially flat background with matter and dark energy, the corresponding dimensionless Hubble rate is
\begin{equation}
E(z)\equiv \frac{H(z)}{H_0}
=\left[\Omega_{\mathrm M}(1+z)^3+(1-\Omega_{\mathrm M})f_{\mathrm{DE}}(z)\right]^{1/2},
\end{equation}
from which we compute the dimensionless comoving distance
\begin{equation}
D_c(z)=\int_0^z \frac{dz'}{E(z')} \, .
\end{equation}
The luminosity and angular-diameter distances then follow as
\begin{equation}
D_L(z)=\frac{c(1+z)}{H_0}D_c(z), 
\qquad
D_A(z)=\frac{c}{H_0(1+z)}D_c(z).
\end{equation}

\subsubsection{Observables used in the iterative search and final evaluation}

These background quantities are mapped to the observables entering the likelihood analysis. For Type Ia supernovae we use the distance modulus,
\begin{equation}
\mu(z)=5\log_{10}\!\left(\frac{D_L(z)}{\mathrm{Mpc}}\right)+25.
\end{equation}
For the weakly model-dependent transverse BAO compilation used in the iterative search, we use the angular BAO scale,
\begin{equation}
\theta_{\mathrm{BAO}}(z)=\frac{r_d}{(1+z)D_A(z)},
\end{equation}
where \(r_d\) is the comoving sound horizon at the drag epoch. To incorporate early-universe information during the search stage without evaluating the full CMB likelihood, we compress the CMB dependence into the standard shift variables
\begin{equation}
R\equiv \sqrt{\Omega_{\mathrm M}}\,D_c(z_*),
\qquad
\ell_A \equiv \pi \frac{D_A(z_*)}{r_s(z_*)},
\end{equation}
where \(r_s(z_*)\) is the comoving sound horizon at photon decoupling and \(z_*\) is the corresponding redshift.

The iterative stage of the pipeline, therefore, uses only the reduced combination \texttt{Pantheon+}+\texttt{2D BAO}+\(\{R,\ell_A\}\), which substantially accelerates candidate screening and refinement. After the iterative search is complete, the selected parameterizations are re-evaluated using the full dataset combinations described in Section~\ref{sec:datasets}, including the complete \texttt{Planck 2018} likelihood and, where applicable, the full BAO measurements such as \texttt{DESI-DR2}. All reported cosmological constraints and model-comparison results are derived from this final full-likelihood analysis.

\subsection{Iterative search objective}
\label{sec:eval}

Candidate equations are ranked during the iterative loop using a search-stage objective defined on the reduced data combination described above. This objective is used only for proposal selection and refinement during the search stage, and not for the final cosmological constraints reported in the paper.

For each proposed equation of state, we compute the residual vector
\begin{equation}
\Delta = x_{\mathrm{obs}}-x_{\mathrm{model}},
\end{equation}
where \(x\) denotes the corresponding observable vector for each dataset. For datasets with an associated covariance matrix, the contribution to the search objective is written in the standard form
\begin{equation}
\chi^2 = \Delta^{\mathrm T} C^{-1}\Delta.
\end{equation}
This treatment is used for the SNIa compilation, the correlated subset of the \texttt{2D BAO} data, and the compressed CMB shift variables \((R,\ell_A)\). In addition, the \texttt{2D BAO} compilation includes three independent transverse BAO measurements for which no non-trivial covariance is provided. Their contribution is therefore added separately as
\begin{equation}
\chi^2_{\mathrm{BAO,ind}}=
\sum_i \frac{\left(\theta_{\mathrm{model},i}-\theta_{\mathrm{obs},i}\right)^2}{\sigma_i^2},
\end{equation}
where \(\sigma_i\) denotes the reported uncertainty of each independent measurement. The full BAO contribution entering the search objective is then
\begin{equation}
\chi^2_{\mathrm{BAO}}=
\chi^2_{\mathrm{BAO,cov}}+\chi^2_{\mathrm{BAO,ind}}.
\end{equation}

The data-based search score is
\begin{equation}
S_{\mathrm{data}}=
\chi^2_{\mathrm{SNIa}}+
\chi^2_{\mathrm{BAO}}+
\chi^2_{\mathrm{CMB}}.
\end{equation}
At this stage, the framework minimizes \(S_{\mathrm{data}}\) with respect to the equation-of-state parameters together with \(H_0\) and \(\Omega_{\mathrm M}\). To combine statistical fit with the language-model critic, we define a heuristic ranking score \(S_{\mathrm{rank}}\) by rescaling \(S_{\mathrm{data}}\) with the critic score \(s_{\mathrm{crit}}\in(0,1]\), so that proposals judged to be weakly motivated or poorly implemented are penalized even when their data fit is competitive. This ranking score is used only within the iterative search and has no direct statistical interpretation. All cosmological constraints and model-comparison results reported in the paper are derived instead from the subsequent full-likelihood analysis.

\subsection{Post-selection cosmological inference}
\label{sec:postselection}

After completing the iterative search, we selected the 25 highest-ranked candidate equations of state for full cosmological inference. Each selected parameterization was then re-analysed using the complete likelihood pipeline, including the full Planck 2018 CMB likelihood rather than the compressed shift-parameter approximation used during the search stage.

For this post-selection analysis, we developed an interface between a modified version of the Boltzmann solver \texttt{CLASS}~\cite{Blas:2011rf} and the Monte Carlo sampler \texttt{MontePython}~\cite{Audren:2012wb,Brinckmann:2018cvx}. This framework allows an arbitrary input parameterization of \(w(a)\) to be passed directly to the background and perturbation solver, after which the Einstein--Boltzmann system is evolved and the corresponding cosmological parameters are sampled with Markov chain Monte Carlo (MCMC).
Convergence of all chains was assessed using the Gelman--Rubin diagnostic~\cite{Gelman:1992zz}, requiring \(R-1\leq 10^{-2}\) in every run.

We assume flat priors on all sampled parameters, in addition to the six standard \(\Lambda\)CDM parameters: \(\omega_{\rm b}\equiv \Omega_{\rm b}h^2\), \(\omega_{\rm cdm}\equiv \Omega_{\rm c}h^2\), \(100\theta_{\rm s}\), \(\ln(10^{10}A_{\rm s})\), \(n_{\rm s}\), and \(\tau_{\rm reio}\). The prior ranges for the baseline cosmological parameters are
\begin{align}
\omega_{\rm b} &\in [0,1], &
\omega_{\rm cdm} &\in [0,1], &
100\theta_{\rm s} &\in [0.5,2.0], \nonumber\\
\ln(10^{10}A_{\rm s}) &\in [1,5], &
n_{\rm s} &\in [0.1,2.0], &
\tau_{\rm reio} &\in [0.004,0.8].
\end{align}
For the standard dark-energy parameterizations (CPL, BA, EXP and JBP), we adopt flat priors \(w_0\in[-3,1]\) and \(w_a\in[-3,2]\). For the AI-generated parameterizations, we use the broader ranges \(w_0\in[-3,3]\) and \(w_a\in[-3,3]\), so their evidences include a conservative Occam penalty
relative to narrower standard-model priors and the post-selection inference does not artificially constrain their phenomenological freedom relative to the baseline models.

Posterior constraints and marginalized confidence regions were obtained with \texttt{GetDist}\footnote{\url{https://github.com/cmbant/getdist}}. We perform the full analysis for each combination of one CMB dataset, one BAO dataset, and one SNIa compilation described in Section~\ref{sec:datasets}, yielding eight dataset combinations in total. The resulting cosmological parameter constraints are reported in the Appendix~\ref{appendix-A}. For model comparison, we evaluate the minimum \(\Delta \chi^2\), the logarithm of the Bayesian evidence, and the corresponding evidence-based significance relative to \(\Lambda\)CDM, \(\sigma\). These results are summarized in Table~\ref{tab:combined_results_performance}.

\subsubsection{Bayesian model comparison}

We compare each dynamical dark-energy (DDE) parameterization with \(\Lambda\)CDM using the Bayesian evidence, which quantifies model performance while accounting for the effective prior volume. For a dataset \(\mathbf{x}\), the Bayes factor between a DDE model \(\mathcal{M}_i\) and \(\Lambda\)CDM is
\begin{equation}
\mathcal{B}_{\Lambda i}
=
\frac{p(\mathbf{x}\mid \Lambda{\rm CDM})}
{p(\mathbf{x}\mid \mathcal{M}_i)}
=
\frac{\displaystyle \int d\boldsymbol{\theta}_{\Lambda} \,
\pi(\boldsymbol{\theta}_{\Lambda}\mid \Lambda{\rm CDM})\,
\mathcal{L}(\mathbf{x}\mid \boldsymbol{\theta}_{\Lambda},\Lambda{\rm CDM})}
{\displaystyle \int d\boldsymbol{\theta}_i \,
\pi(\boldsymbol{\theta}_i\mid \mathcal{M}_i)\,
\mathcal{L}(\mathbf{x}\mid \boldsymbol{\theta}_i,\mathcal{M}_i)} \, .
\label{eq:bayesian_evidence}
\end{equation}
where \(\pi(\boldsymbol{\theta}\mid\mathcal{M})\) denotes the prior distribution and \(\mathcal{L}(\mathbf{x}\mid\boldsymbol{\theta},\mathcal{M})\) the likelihood. In this framework, an extended model may still be disfavoured even when it improves the best fit, if that improvement does not compensate for the larger parameter volume explored.

We report model-comparison results in terms of $\ln \mathcal{B}$, interpreted using the scale proposed by Kass and Raftery~\cite{Kass:1995loi}. Specifically, $\lvert \ln \mathcal{B} \rvert < 1$ corresponds to weak evidence, $1 \leq \lvert \ln \mathcal{B} \rvert < 3$ to positive evidence, $3 \leq \lvert \ln \mathcal{B} \rvert < 5$ to strong evidence, and $\lvert \ln \mathcal{B} \rvert \geq 5$ to very strong evidence.

Bayes factors are estimated from the posterior chains with the \texttt{MCEvidence} package\footnote{\url{https://github.com/yabebalFantaye/MCEvidence}.}~\cite{Heavens:2017hkr,Heavens:2017afc}. We also report an evidence-based significance in units of \(\sigma\), obtained from \(\mathcal{B}_{\Lambda i}\) using the conservative conversion adapted from~\cite{Trotta:2008qt}. In this procedure, the Bayes factor is first mapped to a conservative bound on the corresponding \(p\)-value, which is then translated into an equivalent Gaussian significance through the inverse cumulative distribution of the standard normal. 
The quoted \(\sigma\) values are used only as a descriptive monotonic
re-expression of the Bayes factors under the adopted convention; they should
not be interpreted as frequentist detection significances or as global
post-selection significances~\cite{Ong:2026tta}.

\backmatter
\bmhead{Data availability}

This study uses publicly available cosmological datasets. The Type Ia supernova compilations analysed in this work are \texttt{Pantheon+}~\cite{Brout:2022vxf}, \texttt{Union3}~\cite{Rubin:2023jdq}, and \texttt{DESY5}~\cite{DES:2024jxu}. The CMB data are from the \texttt{Planck 2018} legacy release~\cite{Planck:2018vyg,Planck:2019nip,Planck:2018lbu}. The BAO data are taken from DESI DR2~\cite{DESI:2025zgx,DESI:2025zpo} and from the weakly model-dependent transverse BAO compilation described in Refs.~\cite{Menote22,deCarvalho18,deCarvalho21,Avila25,Sabogal:2025qhz}. All data used in this study are available from the corresponding public data releases and cited references. The derived data underlying the main figures are available from the corresponding author upon reasonable request.


\bmhead{Code availability}

The discovery pipeline is available at
\url{https://iadev.cbpf.br/labia/cosmoai}. The repository contains the code used for AI-assisted candidate generation, literature retrieval, critic-based evaluation, scoring and iterative refinement, together with the full set of prompts. Full-likelihood inference used a modified version of \texttt{CLASS} interfaced with \texttt{MontePython}.

\bmhead{Author contributions}

C.R.B. conceived the study, developed the real-data methodology and contributed to the code implementation. B.M.F. contributed to the code development and analysis. M.S. performed the full-likelihood analysis. P.D. analysed the mock data. G.S. developed the retrieval-augmented generation system and contributed to the implementation. A.B. analysed the data and the dark-energy equation of state. All authors discussed the results and contributed to the manuscript.

\bmhead{Acknowledgments}
CRB acknowledges the financial support from CNPq (316072/2021-4) and from
FAPERJ (grants 201.456/2022 and 210.330/2022) and the FINEP contract 01.22.0505.00 (ref. 1891/22) and 01.25.0215.00 (ref. 1021/24).
The authors made use of Sci-Mind servers machines developed by the CBPF AI LAB team and would like to thank P. Russano and M. Portes de Albuquerque for all the support in infrastructure matters. This work was inspired by Shojaee et al. \citep{llmsr}. M.A.S. acknowledges support from CAPES and expresses gratitude to the Observatório Nacional for its hospitality during the early stages of this work. M.A.S. also acknowledges support during the final stages of this work from the University of Trento and the Provincia Autonoma di Trento (PAT, Autonomous Province of Trento). 
AB acknowledges a CNPq fellowship.

\begin{appendices}


\section{Full posterior constraints for all model and dataset combinations}\label{appendix-A}

For completeness, Table~\ref{table:combined_colored} lists the full marginalized posterior constraints on both the baseline cosmological parameters and the dark-energy equation-of-state parameters, for all models and dataset combinations considered in the post-selection MCMC analysis.

\begin{sidewaystable}[htpb!]
\centering
\caption{Marginalized cosmological parameter constraints (68\% credible intervals) for the standard dark-energy parameterizations and the two
AI-generated parameterizations, shown for the full set of dataset combinations considered in the post-selection analysis. The left and right blocks
correspond to the DESI-DR2 BAO + Planck 2018 and weakly model-dependent 2D BAO + Planck 2018 baselines, respectively, supplemented by no
SNIa dataset (Base), Pantheon+ (+PP, bold), Union3 (+Union3, gray), or DESY5 (+DESY5, cyan).}
\label{table:combined_colored}
\renewcommand{\arraystretch}{1.55}
\resizebox{\linewidth}{!}{

\newcommand{\Base}[1]{\textcolor{black}{#1}}
\newcommand{\PP}[1]{\textcolor{black}{\bm{#1}}}
\newcommand{\Union}[1]{\textcolor{gray}{#1}}
\newcommand{\DESY}[1]{\textcolor{blue!70!cyan!80}{#1}}

\begin{tabular}{l|cccccc|cccccc}
\hline
& \multicolumn{6}{c|}{\textbf{DESI DR2 BAO $+$ Planck 2018}} & \multicolumn{6}{c}{\textbf{Weakly-model-dependent 2D BAO $+$ Planck 2018}} \\
\cline{2-13}
\textbf{Parameter}
& \textbf{CPL} & \textbf{JBP} & \textbf{EXP} & \textbf{BA} & \textbf{AI 1}
& \textbf{AI 2} & \textbf{CPL} & \textbf{JBP} & \textbf{EXP} & \textbf{BA} & \textbf{AI 1} & \textbf{AI 2} \\
\hline\hline

& \makecell{Base\\\textbf{+PP}\\\Union{+Union3}\\\DESY{+DESY5}}
& \makecell{Base\\\textbf{+PP}\\\Union{+Union3}\\\DESY{+DESY5}}
& \makecell{Base\\\textbf{+PP}\\\Union{+Union3}\\\DESY{+DESY5}}
& \makecell{Base\\\textbf{+PP}\\\Union{+Union3}\\\DESY{+DESY5}}
& \makecell{Base\\\textbf{+PP}\\\Union{+Union3}\\\DESY{+DESY5}}
& \makecell{Base\\\textbf{+PP}\\\Union{+Union3}\\\DESY{+DESY5}}
& \makecell{Base\\\textbf{+PP}\\\Union{+Union3}\\\DESY{+DESY5}}
& \makecell{Base\\\textbf{+PP}\\\Union{+Union3}\\\DESY{+DESY5}}
& \makecell{Base\\\textbf{+PP}\\\Union{+Union3}\\\DESY{+DESY5}}
& \makecell{Base\\\textbf{+PP}\\\Union{+Union3}\\\DESY{+DESY5}}
& \makecell{Base\\\textbf{+PP}\\\Union{+Union3}\\\DESY{+DESY5}} & \makecell{Base\\\textbf{+PP}\\\Union{+Union3}\\\DESY{+DESY5}}
\\

\hline\hline

$10^{-2}\omega_{b}$ &
\makecell{\Base{$2.239\pm 0.013$}\\\PP{$2.245\pm 0.013$}\\\Union{$2.243\pm 0.014$}\\\DESY{$2.241\pm 0.014$}\vspace{0.2cm}} &
\makecell{\Base{$2.248\pm 0.014$}\\\PP{$2.248\pm 0.013$}\\\Union{$2.248\pm 0.013$}\\\DESY{$2.248\pm 0.013$}\vspace{0.2cm}} &
\makecell{\Base{$2.239\pm 0.014$}\\\PP{$2.244\pm 0.013$}\\\Union{$2.241\pm 0.014$}\\\DESY{$2.242\pm 0.013$}\vspace{0.2cm}} &
\makecell{\Base{$2.239\pm 0.014$}\\\PP{$2.245\pm 0.013$}\\\Union{$2.242\pm 0.013$}\\\DESY{$2.242\pm 0.013$}\vspace{0.2cm}} &
\makecell{\Base{$2.243\pm 0.013$}\\\PP{$2.242\pm 0.013$}\\\Union{$2.243\pm 0.013$}\\\DESY{$2.243\pm 0.013$}\vspace{0.2cm}} &
\makecell{\Base{$2.244\pm 0.013$}\\\PP{$2.247\pm 0.013$}\\\Union{$2.245\pm 0.013$}\\\DESY{$2.244\pm 0.014$}\vspace{0.2cm}} &
\makecell{\Base{$2.241\pm 0.014$}\\\PP{$2.241^{+0.016}_{-0.014}$}\\\Union{$2.240^{+0.013}_{-0.015}$}\\\DESY{$2.240\pm 0.015$}\vspace{0.2cm}} &
\makecell{\Base{$2.238\pm 0.014$}\\\PP{$2.240\pm 0.014$}\\\Union{$2.241\pm 0.014$}\\\DESY{$2.241\pm 0.014$}\vspace{0.2cm}} &
\makecell{\Base{$2.239\pm 0.015$}\\\PP{$2.239\pm 0.014$}\\\Union{$2.238\pm 0.015$}\\\DESY{$2.238\pm 0.014$}\vspace{0.2cm}} &
\makecell{\Base{$2.239\pm 0.014$}\\\PP{$2.240\pm 0.014$}\\\Union{$2.240\pm 0.014$}\\\DESY{$2.240\pm 0.014$}\vspace{0.2cm}} &
\makecell{\Base{$2.242\pm 0.014$}\\\PP{$2.243\pm 0.014$}\\\Union{$2.242\pm 0.016$}\\\DESY{$2.241\pm 0.014$}\vspace{0.2cm}} &
\makecell{\Base{$2.238\pm 0.014$}\\\PP{$2.241\pm 0.015$}\\\Union{$2.239\pm 0.015$}\\\DESY{$2.239\pm 0.014$}\vspace{0.2cm}} \\

$\omega_{\rm cdm}$ &
\makecell{\Base{$0.11972\pm 0.00090$}\\\PP{$0.11900\pm 0.00085$}\\\Union{$0.11938\pm 0.00090$}\\\DESY{$0.11939\pm 0.00088$}\vspace{0.2cm}} &
\makecell{\Base{$0.11850\pm 0.00085$}\\\PP{$0.11848\pm 0.00083$}\\\Union{$0.11858\pm 0.00080$}\\\DESY{$0.11864\pm 0.00082$}\vspace{0.2cm}} &
\makecell{\Base{$0.11981\pm 0.00093$}\\\PP{$0.11910\pm 0.00091$}\\\Union{$0.11945\pm 0.00089$}\\\DESY{$0.11944\pm 0.00084$}\vspace{0.2cm}} &
\makecell{\Base{$0.11972\pm 0.00095$}\\\PP{$0.11902\pm 0.00087$}\\\Union{$0.11949\pm 0.00088$}\\\DESY{$0.11936\pm 0.00085$}\vspace{0.2cm}} &
\makecell{\Base{$0.11919\pm 0.00068$}\\\PP{$0.11934^{+0.00066}_{-0.00081}$}\\\Union{$0.11951\pm 0.00081$}\\\DESY{$0.11935^{+0.00074}_{-0.00088}$}\vspace{0.2cm}} &
\makecell{\Base{$0.11903\pm 0.00083$}\\\PP{$0.11871\pm 0.00085$}\\\Union{$0.11898\pm 0.00087$}\\\DESY{$0.11911\pm 0.00086$}\vspace{0.2cm}} &
\makecell{\Base{$0.1198\pm 0.0012$}\\\PP{$0.1196\pm 0.0011$}\\\Union{$0.1197\pm 0.0011$}\\\DESY{$0.1198\pm 0.0011$}\vspace{0.2cm}} &
\makecell{\Base{$0.1199\pm 0.0011$}\\\PP{$0.1196\pm 0.0011$}\\\Union{$0.1196\pm 0.0011$}\\\DESY{$0.1196\pm 0.0011$}\vspace{0.2cm}} &
\makecell{\Base{$0.1198\pm 0.0012$}\\\PP{$0.1198\pm 0.0011$}\\\Union{$0.1199\pm 0.0012$}\\\DESY{$0.1199\pm 0.0011$}\vspace{0.2cm}} &
\makecell{\Base{$0.1198\pm 0.0012$}\\\PP{$0.1197\pm 0.0011$}\\\Union{$0.1197\pm 0.0011$}\\\DESY{$0.1198\pm 0.0011$}\vspace{0.2cm}} &
\makecell{\Base{$0.1194\pm 0.0011$}\\\PP{$0.11935\pm 0.0011$}\\\Union{$0.1194\pm 0.0011$}\\\DESY{$0.1195\pm 0.0011$}\vspace{0.2cm}} &
\makecell{\Base{$0.1199\pm 0.0011$}\\\PP{$0.1196\pm 0.0011$}\\\Union{$0.1198\pm 0.0011$}\\\DESY{$0.1198\pm 0.0011$}\vspace{0.2cm}} \\

$100\,\theta_s$ &
\makecell{\Base{$1.04191\pm 0.00028$}\\\PP{$1.04197\pm 0.00028$}\\\Union{$1.04194\pm 0.00028$}\\\DESY{$1.04194\pm 0.00029$}\vspace{0.2cm}} &
\makecell{\Base{$1.04203\pm 0.00028$}\\\PP{$1.04204\pm 0.00027$}\\\Union{$1.04204\pm 0.00027$}\\\DESY{$1.04204\pm 0.00029$}\vspace{0.2cm}} &
\makecell{\Base{$1.04188\pm 0.00029$}\\\PP{$1.04196\pm 0.00028$}\\\Union{$1.04192\pm 0.00029$}\\\DESY{$1.04194\pm 0.00028$}\vspace{0.2cm}} &
\makecell{\Base{$1.04190\pm 0.00029$}\\\PP{$1.04198\pm 0.00027$}\\\Union{$1.04194\pm 0.00029$}\\\DESY{$1.04196\pm 0.00027$}\vspace{0.2cm}} &
\makecell{\Base{$1.04197\pm 0.00027$}\\\PP{$1.04193\pm 0.00028$}\\\Union{$1.04192\pm 0.00027$}\\\DESY{$1.04195\pm 0.00027$}\vspace{0.2cm}} &
\makecell{\Base{$1.04198\pm 0.00027$}\\\PP{$1.04201\pm 0.00028$}\\\Union{$1.04197\pm 0.00028$}\\\DESY{$1.04198\pm 0.00027$}\vspace{0.2cm}} &
\makecell{\Base{$1.04190\pm 0.00028$}\\\PP{$1.04189\pm 0.00030$}\\\Union{$1.04188\pm 0.00029$}\\\DESY{$1.04189\pm 0.00029$}\vspace{0.2cm}} &
\makecell{\Base{$1.04188\pm 0.00029$}\\\PP{$1.04190\pm 0.00029$}\\\Union{$1.04193\pm 0.00029$}\\\DESY{$1.04191\pm 0.00029$}\vspace{0.2cm}} &
\makecell{\Base{$1.04190\pm 0.00029$}\\\PP{$1.04189\pm 0.00029$}\\\Union{$1.04185\pm 0.00030$}\\\DESY{$1.04187\pm 0.00028$}\vspace{0.2cm}} &
\makecell{\Base{$1.04189\pm 0.00028$}\\\PP{$1.04189\pm 0.00028$}\\\Union{$1.04190\pm 0.00029$}\\\DESY{$1.04189\pm 0.00029$}\vspace{0.2cm}} &
\makecell{\Base{$1.04192\pm 0.00029$}\\\PP{$1.04192\pm 0.00030$}\\\Union{$1.04193\pm 0.00030$}\\\DESY{$1.04190\pm 0.00029$}\vspace{0.2cm}} &
\makecell{\Base{$1.04189\pm 0.00029$}\\\PP{$1.04194\pm 0.00028$}\\\Union{$1.04189\pm 0.00029$}\\\DESY{$1.04188\pm 0.00028$}\vspace{0.2cm}} \\

$\ln(10^{10}A_s)$ &
\makecell{\Base{$3.041\pm 0.014$}\\\PP{$3.046\pm 0.014$}\\\Union{$3.044\pm 0.014$}\\\DESY{$3.042\pm 0.014$}\vspace{0.2cm}} &
\makecell{\Base{$3.049^{+0.014}_{-0.015}$}\\\PP{$3.050\pm 0.015$}\\\Union{$3.050\pm 0.014$}\\\DESY{$3.048\pm 0.015$}\vspace{0.2cm}} &
\makecell{\Base{$3.041\pm 0.014$}\\\PP{$3.045\pm 0.014$}\\\Union{$3.043\pm 0.014$}\\\DESY{$3.044\pm 0.014$}\vspace{0.2cm}} &
\makecell{\Base{$3.042\pm 0.015$}\\\PP{$3.047\pm 0.014$}\\\Union{$3.043\pm 0.014$}\\\DESY{$3.043\pm 0.014$}\vspace{0.2cm}} &
\makecell{\Base{$3.044\pm 0.015$}\\\PP{$3.045\pm 0.014$}\\\Union{$3.041\pm 0.012$}\\\DESY{$3.044\pm 0.015$}\vspace{0.2cm}} &
\makecell{\Base{$3.046\pm 0.014$}\\\PP{$3.048^{+0.013}_{-0.015}$}\\\Union{$3.046\pm 0.015$}\\\DESY{$3.046\pm 0.014$}\vspace{0.2cm}} &
\makecell{\Base{$3.042\pm 0.014$}\\\PP{$3.043\pm 0.015$}\\\Union{$3.040\pm 0.014$}\\\DESY{$3.040\pm 0.015$}\vspace{0.2cm}} &
\makecell{\Base{$3.043\pm 0.014$}\\\PP{$3.044\pm 0.015$}\\\Union{$3.044\pm 0.015$}\\\DESY{$3.044\pm 0.015$}\vspace{0.2cm}} &
\makecell{\Base{$3.041^{+0.013}_{-0.015}$}\\\PP{$3.041\pm 0.014$}\\\Union{$3.040\pm 0.015$}\\\DESY{$3.040\pm 0.014$}\vspace{0.2cm}} &
\makecell{\Base{$3.042\pm 0.015$}\\\PP{$3.042\pm 0.015$}\\\Union{$3.041\pm 0.014$}\\\DESY{$3.041\pm 0.014$}\vspace{0.2cm}} &
\makecell{\Base{$3.042\pm 0.014$}\\\PP{$3.043\pm 0.014$}\\\Union{$3.042\pm 0.015$}\\\DESY{$3.042\pm 0.014$}\vspace{0.2cm}} &
\makecell{\Base{$3.043\pm 0.015$}\\\PP{$3.043\pm 0.013$}\\\Union{$3.042\pm 0.014$}\\\DESY{$3.042\pm 0.014$}\vspace{0.2cm}} \\

$n_s$ &
\makecell{\Base{$0.9661\pm 0.0037$}\\\PP{$0.9678\pm 0.0036$}\\\Union{$0.9667\pm 0.0037$}\\\DESY{$0.9670\pm 0.0037$}\vspace{0.2cm}} &
\makecell{\Base{$0.9689\pm 0.0037$}\\\PP{$0.9692\pm 0.0035$}\\\Union{$0.9688\pm 0.0035$}\\\DESY{$0.9687\pm 0.0035$}\vspace{0.2cm}} &
\makecell{\Base{$0.9657\pm 0.0036$}\\\PP{$0.9673\pm 0.0036$}\\\Union{$0.9667\pm 0.0037$}\\\DESY{$0.9666\pm 0.0036$}\vspace{0.2cm}} &
\makecell{\Base{$0.9661\pm 0.0037$}\\\PP{$0.9675\pm 0.0037$}\\\Union{$0.9664\pm 0.0036$}\\\DESY{$0.9670\pm 0.0035$}\vspace{0.2cm}} &
\makecell{\Base{$0.9671\pm 0.0033$}\\\PP{$0.9666\pm 0.0034$}\\\Union{$0.9665^{+0.0030}_{-0.0034}$}\\\DESY{$0.9672\pm 0.0035$}\vspace{0.2cm}} &
\makecell{\Base{$0.9676\pm 0.0035$}\\\PP{$0.9684\pm 0.0035$}\\\Union{$0.9676\pm 0.0036$}\\\DESY{$0.9675\pm 0.0036$}\vspace{0.2cm}} &
\makecell{\Base{$0.9660\pm 0.0040$}\\\PP{$0.9665\pm 0.0040$}\\\Union{$0.9664\pm 0.0039$}\\\DESY{$0.9662\pm 0.0040$}\vspace{0.2cm}} &
\makecell{\Base{$0.9656\pm 0.0041$}\\\PP{$0.9662\pm 0.0040$}\\\Union{$0.9663\pm 0.0039$}\\\DESY{$0.9664\pm 0.0039$}\vspace{0.2cm}} &
\makecell{\Base{$0.9658\pm 0.0042$}\\\PP{$0.9658\pm 0.0040$}\\\Union{$0.9654\pm 0.0041$}\\\DESY{$0.9656\pm 0.0040$}\vspace{0.2cm}} &
\makecell{\Base{$0.9660\pm 0.0041$}\\\PP{$0.9661\pm 0.0040$}\\\Union{$0.9661\pm 0.0039$}\\\DESY{$0.9657\pm 0.0039$}\vspace{0.2cm}} &
\makecell{\Base{$0.9668\pm 0.0041$}\\\PP{$0.9671\pm 0.0037$}\\\Union{$0.9668\pm 0.0040$}\\\DESY{$0.9663\pm 0.0041$}\vspace{0.2cm}} &
\makecell{\Base{$0.9658\pm 0.0039$}\\\PP{$0.9662\pm 0.0040$}\\\Union{$0.9658\pm 0.0039$}\\\DESY{$0.9658\pm 0.0040$}\vspace{0.2cm}} \\

$\tau_{\rm reio}$ &
\makecell{\Base{$0.0531\pm 0.0071$}\\\PP{$0.0559\pm 0.0073$}\\\Union{$0.0545\pm 0.0073$}\\\DESY{$0.0540\pm 0.0073$}\vspace{0.2cm}} &
\makecell{\Base{$0.0579^{+0.0069}_{-0.0079}$}\\\PP{$0.0584\pm 0.0074$}\\\Union{$0.0579\pm 0.0070$}\\\DESY{$0.0574\pm 0.0074$}\vspace{0.2cm}} &
\makecell{\Base{$0.0528\pm 0.0072$}\\\PP{$0.0555\pm 0.0072$}\\\Union{$0.0543^{+0.0065}_{-0.0073}$}\\\DESY{$0.0544\pm 0.0071$}\vspace{0.2cm}} &
\makecell{\Base{$0.0533\pm 0.0074$}\\\PP{$0.0563\pm 0.0072$}\\\Union{$0.0538\pm 0.0074$}\\\DESY{$0.0542\pm 0.0071$}\vspace{0.2cm}} &
\makecell{\Base{$0.0549\pm 0.0071$}\\\PP{$0.0548\pm 0.0068$}\\\Union{$0.0532\pm 0.0060$}\\\DESY{$0.0547\pm 0.0075$}\vspace{0.2cm}} &
\makecell{\Base{$0.0557\pm 0.0072$}\\\PP{$0.0572^{+0.0066}_{-0.0078}$}\\\Union{$0.0559\pm 0.0076$}\\\DESY{$0.0558\pm 0.0073$}\vspace{0.2cm}} &
\makecell{\Base{$0.0531\pm 0.0073$}\\\PP{$0.0538\pm 0.0076$}\\\Union{$0.0525\pm 0.0075$}\\\DESY{$0.0523\pm 0.0076$}\vspace{0.2cm}} &
\makecell{\Base{$0.0535\pm 0.0074$}\\\PP{$0.0545\pm 0.0075$}\\\Union{$0.0544^{+0.0069}_{-0.0077}$}\\\DESY{$0.0544\pm 0.0075$}\vspace{0.2cm}} &
\makecell{\Base{$0.0527^{+0.0069}_{-0.0079}$}\\\PP{$0.0532\pm 0.0073$}\\\Union{$0.0521\pm 0.0075$}\\\DESY{$0.0524\pm 0.0073$}\vspace{0.2cm}} &
\makecell{\Base{$0.0535\pm 0.0078$}\\\PP{$0.0533^{+0.0070}_{-0.0078}$}\\\Union{$0.0530\pm 0.0075$}\\\DESY{$0.0527\pm 0.0074$}\vspace{0.2cm}} &
\makecell{\Base{$0.0533\pm 0.0073$}\\\PP{$0.0542\pm 0.0071$}\\\Union{$0.0537^{+0.0068}_{-0.0077}$}\\\DESY{$0.0534\pm 0.0074$}\vspace{0.2cm}} &
\makecell{\Base{$0.0535\pm 0.0080$}\\\PP{$0.0543\pm 0.0068$}\\\Union{$0.0531\pm 0.0074$}\\\DESY{$0.0534\pm 0.0073$}\vspace{0.2cm}} \\

\hline

$w_{0}$ &
\makecell{\Base{$-0.45\pm 0.20$}\\\PP{$-0.832\pm 0.054$}\\\Union{$-0.678\pm 0.087$}\\\DESY{$-0.693\pm 0.058$}\vspace{0.2cm}} &
\makecell{\Base{$-0.648^{+0.25}_{-0.086}$}\\\PP{$-0.800\pm 0.076$}\\\Union{$-0.586^{+0.12}_{-0.078}$}\\\DESY{$-0.583^{+0.083}_{-0.062}$}\vspace{0.2cm}} &
\makecell{\Base{$-0.53^{+0.17}_{-0.22}$}\\\PP{$-0.846\pm 0.051$}\\\Union{$-0.712\pm 0.080$}\\\DESY{$-0.723\pm 0.050$}\vspace{0.2cm}} &
\makecell{\Base{$-0.52^{+0.17}_{-0.20}$}\\\PP{$-0.859\pm 0.046$}\\\Union{$-0.715\pm 0.075$}\\\DESY{$-0.741\pm 0.049$}\vspace{0.2cm}} &
\makecell{\Base{$0.20^{+0.12}_{-0.17}$}\\\PP{$0.28^{+0.18}_{-0.24}$}\\\Union{$0.45^{+0.20}_{-0.38}$}\\\DESY{$0.50^{+0.25}_{-0.29}$}\vspace{0.2cm}} &
\makecell{\Base{$-0.40^{+0.23}_{-0.10}$}\\\PP{$-0.817\pm 0.065$}\\\Union{$-0.60\pm 0.11$}\\\DESY{$-0.633\pm 0.070$}\vspace{0.2cm}} &
\makecell{\Base{$-0.85\pm 0.38$}\\\PP{$-0.834\pm 0.062$}\\\Union{$-0.69\pm 0.10$}\\\DESY{$-0.672\pm 0.071$}\vspace{0.2cm}} &
\makecell{\Base{$-1.05^{+0.26}_{-0.38}$}\\\PP{$-0.792\pm 0.088$}\\\Union{$-0.626^{+0.13}_{-0.073}$}\\\DESY{$-0.585^{+0.084}_{-0.049}$}\vspace{0.2cm}} &
\makecell{\Base{$-0.77^{+0.30}_{-0.39}$}\\\PP{$-0.847\pm 0.060$}\\\Union{$-0.693\pm 0.098$}\\\DESY{$-0.699\pm 0.065$}\vspace{0.2cm}} &
\makecell{\Base{$-0.84^{+0.36}_{-0.42}$}\\\PP{$-0.866\pm 0.054$}\\\Union{$-0.733\pm 0.087$}\\\DESY{$-0.723\pm 0.059$}\vspace{0.2cm}} &
\makecell{\Base{$0.57^{+0.23}_{-0.53}$}\\\PP{$0.39^{+0.16}_{-0.32}$}\\\Union{$0.68^{+0.31}_{-0.51}$}\\\DESY{$0.86^{+0.26}_{-0.33}$}\vspace{0.2cm}} &
\makecell{\Base{$-1.00^{+0.37}_{-0.43}$}\\\PP{$-0.840^{+0.071}_{-0.085}$}\\\Union{$-0.62\pm 0.12$}\\\DESY{$-0.614\pm 0.078$}\vspace{0.2cm}} \\

$w_{\rm a}$ &
\makecell{\Base{$-1.66\pm 0.58$}\\\PP{$-0.61\pm 0.20$}\\\Union{$-1.04^{+0.32}_{-0.27}$}\\\DESY{$-1.01^{+0.25}_{-0.22}$}\vspace{0.2cm}} &
\makecell{\Base{$< -1.62$}\\\PP{$-1.16\pm 0.45$}\\\Union{$-2.17^{+0.30}_{-0.73}$}\\\DESY{$-2.22^{+0.36}_{-0.53}$}\vspace{0.2cm}} &
\makecell{\Base{$-1.15^{+0.52}_{-0.38}$}\\\PP{$-0.47^{+0.18}_{-0.15}$}\\\Union{$-0.76^{+0.24}_{-0.20}$}\\\DESY{$-0.74^{+0.17}_{-0.15}$}\vspace{0.2cm}} &
\makecell{\Base{$-0.79^{+0.31}_{-0.26}$}\\\PP{$-0.293^{+0.10}_{-0.091}$}\\\Union{$-0.52\pm 0.14$}\\\DESY{$-0.476^{+0.11}_{-0.099}$}\vspace{0.2cm}} &
\makecell{\Base{$-0.33\pm 0.21$}\\\PP{$-0.47^{+0.35}_{-0.18}$}\\\Union{$-0.50^{+0.39}_{-0.27}$}\\\DESY{$-0.385^{+0.22}_{-0.082}$}\vspace{0.2cm}} &
\makecell{\Base{$< -1.99$}\\\PP{$-0.80\pm 0.29$}\\\Union{$-1.53^{+0.45}_{-0.40}$}\\\DESY{$-1.47^{+0.34}_{-0.31}$}\vspace{0.2cm}} &
\makecell{\Base{$-0.7^{+1.2}_{-1.0}$}\\\PP{$-0.71\pm 0.26$}\\\Union{$-1.13\pm 0.38$}\\\DESY{$-1.20^{+0.33}_{-0.28}$}\vspace{0.2cm}} &
\makecell{\Base{$> -1.06$}\\\PP{$-1.38\pm 0.57$}\\\Union{$< -1.96$}\\\DESY{$< -2.21$}\vspace{0.2cm}} &
\makecell{\Base{$-0.77^{+0.92}_{-0.60}$}\\\PP{$-0.55^{+0.23}_{-0.20}$}\\\Union{$-0.94^{+0.35}_{-0.28}$}\\\DESY{$-0.92^{+0.26}_{-0.21}$}\vspace{0.2cm}} &
\makecell{\Base{$-0.41^{+0.65}_{-0.53}$}\\\PP{$-0.35^{+0.14}_{-0.12}$}\\\Union{$-0.56^{+0.20}_{-0.17}$}\\\DESY{$-0.59^{+0.16}_{-0.14}$}\vspace{0.2cm}} &
\makecell{\Base{$-0.208^{+0.16}_{-0.090}$}\\\PP{$-0.29^{+0.21}_{-0.16}$}\\\Union{$-0.284^{+0.15}_{-0.026}$}\\\DESY{$-0.229^{+0.071}_{-0.030}$}\vspace{0.2cm}} &
\makecell{\Base{$-0.4^{+1.5}_{-1.4}$}\\\PP{$-0.80^{+0.43}_{-0.34}$}\\\Union{$-1.65\pm 0.51$}\\\DESY{$-1.70\pm 0.39$}\vspace{0.2cm}} \\

$H_0 \, [\mathrm{km/s/Mpc}]$ &
\makecell{\Base{$64.0^{+1.7}_{-2.0}$}\\\PP{$67.58\pm 0.61$}\\\Union{$66.12\pm 0.83$}\\\DESY{$66.35\pm 0.57$}\vspace{0.2cm}} &
\makecell{\Base{$66.5^{+1.0}_{-2.0}$}\\\PP{$67.63\pm 0.61$}\\\Union{$66.07\pm 0.81$}\\\DESY{$66.16\pm 0.56$}\vspace{0.2cm}} &
\makecell{\Base{$64.4\pm 1.9$}\\\PP{$67.64\pm 0.58$}\\\Union{$66.24\pm 0.82$}\\\DESY{$66.39\pm 0.55$}\vspace{0.2cm}} &
\makecell{\Base{$64.1\pm 1.9$}\\\PP{$67.61\pm 0.59$}\\\Union{$66.15\pm 0.80$}\\\DESY{$66.42\pm 0.55$}\vspace{0.2cm}} &
\makecell{\Base{$68.31\pm 0.52$}\\\PP{$67.64^{+0.66}_{-0.59}$}\\\Union{$66.80^{+0.77}_{-0.69}$}\\\DESY{$66.69\pm 0.56$}\vspace{0.2cm}} &
\makecell{\Base{$64.1^{+1.1}_{-2.0}$}\\\PP{$67.59\pm 0.60$}\\\Union{$65.83\pm 0.90$}\\\DESY{$66.21\pm 0.55$}\vspace{0.2cm}} &
\makecell{\Base{$68.6^{+3.3}_{-3.9}$}\\\PP{$68.03\pm 0.64$}\\\Union{$66.86\pm 0.86$}\\\DESY{$66.81\pm 0.61$}\vspace{0.2cm}} &
\makecell{\Base{$70.3\pm 2.4$}\\\PP{$67.87\pm 0.64$}\\\Union{$66.73\pm 0.87$}\\\DESY{$66.46\pm 0.59$}\vspace{0.2cm}} &
\makecell{\Base{$67.7\pm 3.2$}\\\PP{$68.04\pm 0.67$}\\\Union{$66.85\pm 0.87$}\\\DESY{$66.87\pm 0.59$}\vspace{0.2cm}} &
\makecell{\Base{$68.2\pm 4.0$}\\\PP{$68.08\pm 0.64$}\\\Union{$66.96\pm 0.84$}\\\DESY{$66.91\pm 0.60$}\vspace{0.2cm}} &
\makecell{\Base{$67.9^{+1.3}_{-0.91}$}\\\PP{$68.02\pm 0.56$}\\\Union{$67.23\pm 0.80$}\\\DESY{$67.05\pm 0.59$}\vspace{0.2cm}} &
\makecell{\Base{$69.9\pm 3.2$}\\\PP{$67.85\pm 0.65$}\\\Union{$66.60\pm 0.89$}\\\DESY{$66.63\pm 0.60$}\vspace{0.2cm}} \\

$\sigma_8$ &
\makecell{\Base{$0.783\pm 0.016$}\\\PP{$0.8097\pm 0.0091$}\\\Union{$0.7999\pm 0.0099$}\\\DESY{$0.8016\pm 0.0090$}\vspace{0.2cm}} &
\makecell{\Base{$0.797^{+0.014}_{-0.019}$}\\\PP{$0.8067\pm 0.0092$}\\\Union{$0.794\pm 0.010$}\\\DESY{$0.7950\pm 0.0089$}\vspace{0.2cm}} &
\makecell{\Base{$0.788^{+0.018}_{-0.015}$}\\\PP{$0.8110\pm 0.0093$}\\\Union{$0.802\pm 0.010$}\\\DESY{$0.8029\pm 0.0085$}\vspace{0.2cm}} &
\makecell{\Base{$0.784\pm 0.016$}\\\PP{$0.8104\pm 0.0094$}\\\Union{$0.800\pm 0.010$}\\\DESY{$0.8017\pm 0.0088$}\vspace{0.2cm}} &
\makecell{\Base{$0.8170\pm 0.0077$}\\\PP{$0.8128\pm 0.0086$}\\\Union{$0.8053\pm 0.0093$}\\\DESY{$0.8046^{+0.0078}_{-0.0089}$}\vspace{0.2cm}} &
\makecell{\Base{$0.780^{+0.013}_{-0.018}$}\\\PP{$0.8079\pm 0.0093$}\\\Union{$0.795\pm 0.011$}\\\DESY{$0.7987\pm 0.0089$}\vspace{0.2cm}} &
\makecell{\Base{$0.822^{+0.030}_{-0.034}$}\\\PP{$0.817\pm 0.010$}\\\Union{$0.808\pm 0.011$}\\\DESY{$0.8082\pm 0.0096$}\vspace{0.2cm}} &
\makecell{\Base{$0.836\pm 0.022$}\\\PP{$0.815\pm 0.010$}\\\Union{$0.805\pm 0.011$}\\\DESY{$0.8027\pm 0.0095$}\vspace{0.2cm}} &
\makecell{\Base{$0.816\pm 0.029$}\\\PP{$0.818\pm 0.010$}\\\Union{$0.810\pm 0.011$}\\\DESY{$0.8099\pm 0.0095$}\vspace{0.2cm}} &
\makecell{\Base{$0.819\pm 0.035$}\\\PP{$0.818\pm 0.010$}\\\Union{$0.809\pm 0.011$}\\\DESY{$0.8097\pm 0.0098$}\vspace{0.2cm}} &
\makecell{\Base{$0.814\pm 0.012$}\\\PP{$0.8156^{+0.0081}_{-0.0096}$}\\\Union{$0.810\pm 0.011$}\\\DESY{$0.8093\pm 0.0093$}\vspace{0.2cm}} &
\makecell{\Base{$0.833\pm 0.028$}\\\PP{$0.814\pm 0.011$}\\\Union{$0.806\pm 0.011$}\\\DESY{$0.807\pm 0.010$}\vspace{0.2cm}} \\

$\Omega_{\rm m}$ &
\makecell{\Base{$0.350\pm 0.020$}\\\PP{$0.3112\pm 0.0058$}\\\Union{$0.3258\pm 0.0086$}\\\DESY{$0.3237\pm 0.0059$}\vspace{0.2cm}} &
\makecell{\Base{$0.320^{+0.019}_{-0.011}$}\\\PP{$0.3096\pm 0.0057$}\\\Union{$0.3245\pm 0.0080$}\\\DESY{$0.3237\pm 0.0056$}\vspace{0.2cm}} &
\makecell{\Base{$0.345^{+0.018}_{-0.023}$}\\\PP{$0.3108\pm 0.0056$}\\\Union{$0.3248\pm 0.0083$}\\\DESY{$0.3233\pm 0.0057$}\vspace{0.2cm}} &
\makecell{\Base{$0.348^{+0.020}_{-0.023}$}\\\PP{$0.3109\pm 0.0056$}\\\Union{$0.3258\pm 0.0083$}\\\DESY{$0.3228\pm 0.0056$}\vspace{0.2cm}} &
\makecell{\Base{$0.3049\pm 0.0052$}\\\PP{$0.3114^{+0.0055}_{-0.0065}$}\\\Union{$0.3197\pm 0.0079$}\\\DESY{$0.3203\pm 0.0057$}\vspace{0.2cm}} &
\makecell{\Base{$0.347^{+0.021}_{-0.013}$}\\\PP{$0.3104^{+0.0053}_{-0.0060}$}\\\Union{$0.3278\pm 0.0092$}\\\DESY{$0.3243\pm 0.0057$}\vspace{0.2cm}} &
\makecell{\Base{$0.306^{+0.029}_{-0.034}$}\\\PP{$0.3084\pm 0.0060$}\\\Union{$0.3195\pm 0.0085$}\\\DESY{$0.3200\pm 0.0063$}\vspace{0.2cm}} &
\makecell{\Base{$0.290^{+0.018}_{-0.022}$}\\\PP{$0.3097\pm 0.0063$}\\\Union{$0.3202\pm 0.0085$}\\\DESY{$0.3227\pm 0.0062$}\vspace{0.2cm}} &
\makecell{\Base{$0.313^{+0.025}_{-0.033}$}\\\PP{$0.3085\pm 0.0064$}\\\Union{$0.3199\pm 0.0085$}\\\DESY{$0.3195\pm 0.0061$}\vspace{0.2cm}} &
\makecell{\Base{$0.310^{+0.030}_{-0.039}$}\\\PP{$0.3081^{+0.0057}_{-0.0065}$}\\\Union{$0.3185\pm 0.0082$}\\\DESY{$0.3191\pm 0.0061$}\vspace{0.2cm}} &
\makecell{\Base{$0.3096^{+0.0086}_{-0.012}$}\\\PP{$0.3079^{+0.0052}_{-0.0059}$}\\\Union{$0.3154\pm 0.0080$}\\\DESY{$0.3171\pm 0.0061$}\vspace{0.2cm}} &
\makecell{\Base{$0.294^{+0.025}_{-0.031}$}\\\PP{$0.3098\pm 0.0063$}\\\Union{$0.3221\pm 0.0088$}\\\DESY{$0.3216\pm 0.0062$}\vspace{0.2cm}} \\

$r_d \, [\mathrm{Mpc}]$ &
\makecell{\Base{$147.14\pm 0.23$}\\\PP{$147.26\pm 0.21$}\\\Union{$147.19\pm 0.22$}\\\DESY{$147.20\pm 0.22$}} &
\makecell{\Base{$147.37\pm 0.22$}\\\PP{$147.37\pm 0.21$}\\\Union{$147.34\pm 0.21$}\\\DESY{$147.33\pm 0.22$}} &
\makecell{\Base{$147.12\pm 0.23$}\\\PP{$147.25\pm 0.23$}\\\Union{$147.19\pm 0.22$}\\\DESY{$147.18\pm 0.21$}} &
\makecell{\Base{$147.14\pm 0.23$}\\\PP{$147.26\pm 0.21$}\\\Union{$147.17\pm 0.22$}\\\DESY{$147.20\pm 0.21$}} &
\makecell{\Base{$147.23\pm 0.19$}\\\PP{$147.21^{+0.22}_{-0.18}$}\\\Union{$147.15\pm 0.20$}\\\DESY{$147.20\pm 0.20$}} &
\makecell{\Base{$147.26\pm 0.21$}\\\PP{$147.32\pm 0.21$}\\\Union{$147.28\pm 0.22$}\\\DESY{$147.25\pm 0.22$}} &
\makecell{\Base{$147.09\pm 0.26$}\\\PP{$147.14\pm 0.25$}\\\Union{$147.13\pm 0.25$}\\\DESY{$147.12\pm 0.25$}} &
\makecell{\Base{$147.11\pm 0.26$}\\\PP{$147.15\pm 0.25$}\\\Union{$147.16\pm 0.25$}\\\DESY{$147.16\pm 0.24$}} &
\makecell{\Base{$147.11\pm 0.26$}\\\PP{$147.13\pm 0.25$}\\\Union{$147.09\pm 0.26$}\\\DESY{$147.11\pm 0.25$}} &
\makecell{\Base{$147.13\pm 0.26$}\\\PP{$147.13\pm 0.25$}\\\Union{$147.14\pm 0.26$}\\\DESY{$147.10\pm 0.25$}} &
\makecell{\Base{$147.20\pm 0.25$}\\\PP{$147.20\pm 0.24$}\\\Union{$147.19^{+0.28}_{-0.23}$}\\\DESY{$147.18\pm 0.25$}} &
\makecell{\Base{$147.12\pm 0.25$}\\\PP{$147.16\pm 0.25$}\\\Union{$147.12\pm 0.25$}\\\DESY{$147.12\pm 0.25$}} \\

\hline\hline
\end{tabular}
}

\end{sidewaystable}


\section{Blind validation on a mock exotic cosmology}\label{appendix-B}

\begin{figure*}[h!]
    \centering
    \includegraphics[width=1\linewidth]{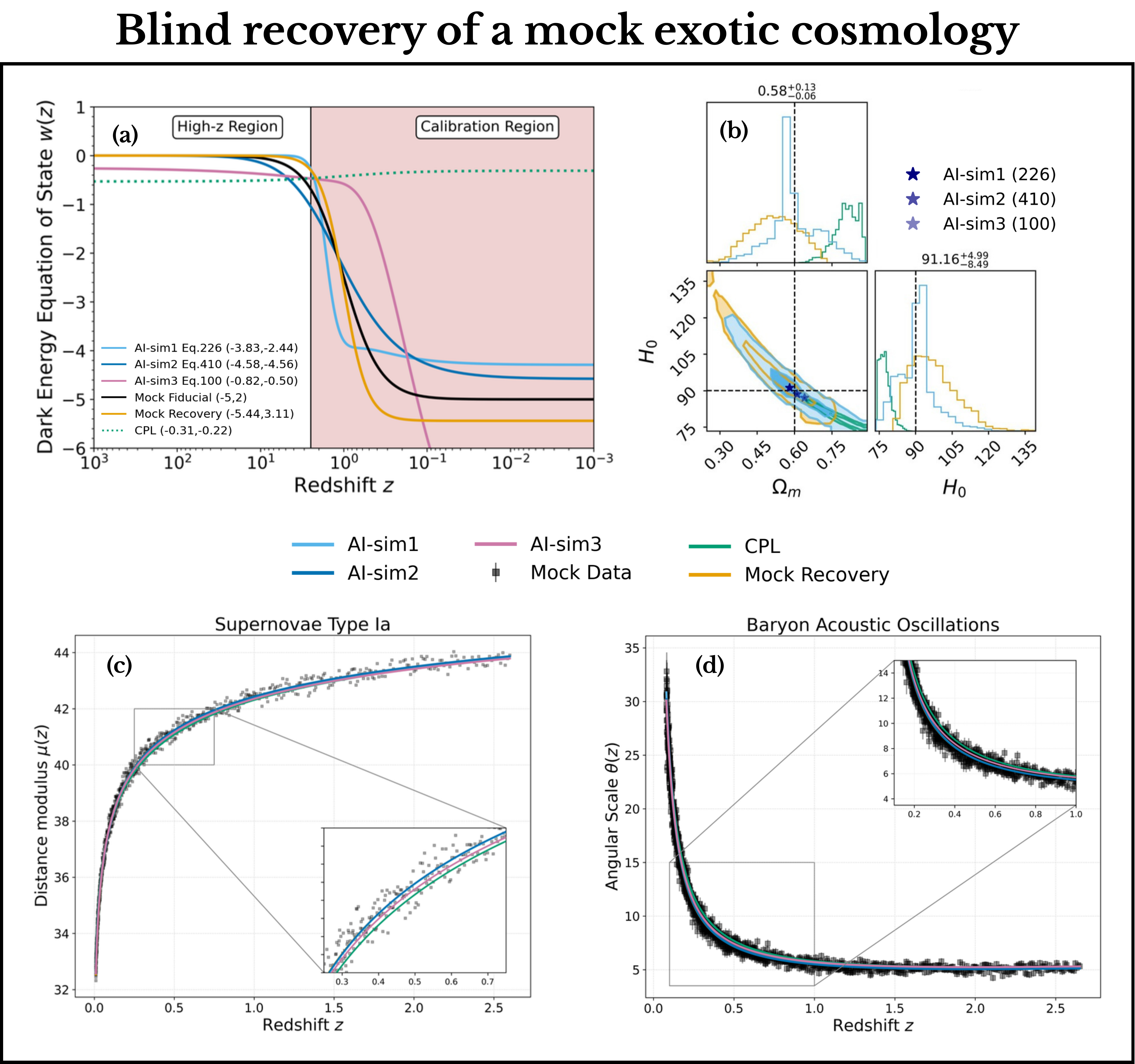}
    \caption{\textbf{Blind recovery of a mock exotic cosmology.} \textbf{a,} Dark-energy equation of state $w(z)$ for the mock fiducial model, its direct recovery from the simulated data, CPL, and the three representative AI-discovered parameterizations. The shaded region marks the redshift range used for calibration. \textbf{b,} Joint and marginalized posteriors in the $(\Omega_m, H_0)$ plane, with dashed lines indicating the fiducial values used to generate the mock cosmology. \textbf{c,} Distance modulus $\mu(z)$ for the mock Type Ia supernova sample. \textbf{d,} Angular BAO scale $\theta(z)$ for the mock BAO sample. In \textbf{c} and \textbf{d}, black points denote the simulated data and insets show zoomed views of the fitted relations.}
   
    \label{fig:mock_cosmo}
\end{figure*}

Figure~\ref{fig:mock_cosmo} shows the blind recovery test for the mock exotic cosmology, comparing the AI-discovered parameterizations with CPL and with the injected fiducial model. Panel~\textbf{a} presents the reconstructed equation of state, $w(z)$, evaluated at the best-fit parameters. The shaded region marks the redshift interval directly covered by the simulated data. In this range, the AI-generated models reproduce the recovered fiducial evolution much more closely than CPL, which fails to capture the sharp low-redshift transition. Notably, AI-sim1 and AI-sim2 approach nearly the same effective behaviour despite having different analytic forms.

Panel~\textbf{b} shows the posterior constraints in the $(\Omega_m,H_0)$ plane. The AI-discovered models recover the injected background cosmology close to the fiducial values, whereas CPL yields a biased solution, reflecting its inability to accommodate the non-standard dark-energy evolution without shifting the background parameters. Panels~\textbf{c} and \textbf{d} show the fits to the mock supernova distance modulus and BAO angular scale. Although all models provide visually reasonable fits, the AI-generated parameterizations yield smaller residuals than CPL and remain nearly indistinguishable from the direct recovery of the injected model over most of the observed redshift range.

%
\section{Reasoning and critique of AI proposed equations}\label{appendix-C}

The candidate models emerge from a long iterative reasoning chain, involving thousands of steps of proposal generation, extended model deliberation, critic-guided feedback, and retrieval of high-performing equations stored in the memory buffer from earlier rounds. Because large language models do not maintain fully uniform conceptual consistency across such extended generations, the intermediate trace may contain tentative physical interpretations, conceptual inconsistencies, or partially conflicting lines of argument. In our framework, these imperfections are not suppressed; they form part of the discovery process and are subsequently exposed, criticized, and often corrected in later exchanges between the generator and the LLM critic and later the human-in-the-loop. To illustrate this process, we present selected final segments of the reasoning chain from which AI 1 and AI 2 emerged through the generator–critic-data interaction. For readability, we have removed residual artefacts from the raw output and lightly edited the presentation style, including the addition of bullet points.

\vspace{.3cm}
\par \textbf{AI 1 reasoning}
\vspace{.2cm}
\begin{mdframed}
\ttfamily
The provided parameterizations (v0-v2) 
yield $w(0) = -1$.
To correct this while maintaining physical consistency ($w(z) > -1$ at high $z$ to avoid unobserved early acceleration), 
I introduce a constant shift $-p_0 p_1$ derived from the existing parameters. This ensures:
\begin{itemize}
    \item $w(0) = -1 - p_0 p_1 < -1$ (satisfies current acceleration),
    \item  $w(\infty) = -1 - p_0 + p_1 - p_0 p_1 > -1$ under $p_1 > \frac{p_0}{1-p_0}$ (prevents early acceleration),
    \item A single crossing of $w(z) = -1$ (transition redshift $z_t < 1$ implied by data).
\end{itemize}
The parameterization uses the standard scaling $x = z^2/(1+z^2)$ for smooth behavior, with a quadratic term to capture the necessary curvature. It is simple (polynomial operations only), uses exactly two parameters, and meets all physical requirements without explicit $z<1$ dependence.

This is implemented as equation v3 with clear documentation of the physics.

\begin{verbatim}
python 
def equation_v3(z: np.ndarray, params: np.ndarray) 
    -> np.ndarray: 
    """Physical parameterization for dynamical 
    dark energy EOS with w(0) < -1 and w(inf) > -1.
    
    The equation of state is:
        w(z) = -1 - p0 * (z^2/(1+z^2)) + p1 * 
        (z^2/(1+z^2))^2 - p0 * p1
    with 0 < p0 < 1 and p1 > p0/(1-p0) 
    (ensuring physical behavior: w(0) < -1 and w(\infty) > -1).
    
    Args:
        z: Redshift array (0 <=  z <=  2.3 for SNIa data).
        params: [p0, p1] (p0 > 0, p1 > p0/(1-p0), p0 < 1).
    
    Returns:
        Array of w(z) values for the given redshifts.
    """
    p0, p1 = params
    x = z**2 / (1 + z**2)
    return -1.0 - p0 * x + p1 * (x ** 2) - p0 * p1
\end{verbatim}
\end{mdframed}
\vspace{0.3cm}
\textbf{AI 1 critique}
\vspace{.2cm}
\begin{mdframed}
    \ttfamily
    Summary: the proposed $w(z) = -1 - p0 x + p1 x^2 - p0 p1$ with $x = z^2/(1+z^2)$ is a simple, two-parameter, bounded-in-x phenomenological ansatz that avoids 
    $w(0)=-1$ by introducing a constant shift. It is usable, but has important caveats.
\subsection*{Physical validity}
    \begin{itemize}
       \item Pros
       \begin{itemize}
          \item $x\in [0,1)$ for $z\in[0,\infty)$, so the functional form is bounded in the chosen variable and smooth.
          \item $w(0)=-1-p0 p1 < -1$ (for $p0,p1>0$) so the model can represent a present-day phantom-like w.
          \item $w(\infty)=-1-p0+p1-p0 p1$; one can choose parameters so $w(\infty)>-1$ (no early acceleration).
          \end{itemize}
         \item Cons / caveats
         \begin{itemize}
          \item The constant shift $-p0 p1$ is an ad hoc algebraic fix (no physical derivation). It is not physically motivated by a field or fluid model; it is purely phenomenological.
         \item Phantom behavior ($w<-1$) and especially crossings of $w=-1$ are theoretically nontrivial: single canonical scalar fields cannot cross $w=-1$ stably, so the model implicitly requires exotic physics (multiple fields, noncanonical kinetic terms or modified gravity). This should be noted when interpreting fits.
         \item Depending on parameters the model can have zero, one or two roots of $w+1=0$ inside $x\in[0,1]$, i.e. it may have multiple crossings of the phantom divide within the observable redshift range. That can be undesirable or unphysical for some theories; the parameter-space region that ensures a single crossing (and $z_t<1$) must be characterized and/or priored.
         \item The physical requirement $w(\infty)>-1$ imposes $p1(1-p0)>p0$ (i.e. $p1 > p0/(1-p0)$ for $p0<1$). Those constraints must be enforced as priors in inference; otherwise unphysical early acceleration can occur.
         \end{itemize}
         \end{itemize}

\subsection*{Mathematical/numerical correctness}
 \begin{itemize} 
\item Implementation is algebraically correct and numerically cheap.
Potential numerical issues:
  \item If p0 or p1 are large (or of opposite sign), w can be large in magnitude; you should bound parameters in inference to avoid extreme/unphysical values and floating-point overflow.
  \item  For very large z, x→1 so direct evaluation is fine; for z near machine precision limits no special pathologies arise here.
  \item  The code assumes params length==2 and z is numeric/array; production code should validate inputs and broadcast correctly.
\item  Analytic properties:
   The crossing condition $w=-1$ reduces to $p1 x^2 - p0 x - p0 p1 =0$, so two real roots always exist for real positive $p0,p1$ (discriminant $>0$). Whether those roots fall inside $[0,1]$ depends on parameters; hence multiple crossings in the physical interval are possible and must be checked.
\end{itemize}
\subsection*{Originality \& creativity}
\begin{itemize}
    \item  Modest originality: the form is a straightforward polynomial-in-a-bounded-mapping of $z (x=z^2/(1+z^2))$, which is common in phenomenological parameterizations.
    \item The trick of shifting the whole curve by -p0 p1 to force $w(0)<-1$ while keeping only two parameters is an ad hoc algebraic maneuver rather than a physically motivated innovation. It is clever in the sense of fixing the immediate bug without adding parameters, but not conceptually new compared to standard parameterizations (e.g. CPL, polynomial-in-a or in x).
    \item Recommendation if novelty is desired: parameterize w0 explicitly ($w(0)=w_0$) plus one shape parameter, or use a two-parameter physically-inspired mapping (e.g. a CPL-like pair w0, wa or a scalar-field motivated form).
\end{itemize}

\subsection*{Clarity of reasoning}
\begin{itemize}
    \item The LLM eventually explains why $w(0)$ must be $<-1$ and how the constant shift fixes that; it also gives the condition for $w(\infty)>-1$. That is useful.
     \item Weaknesses: the earlier narrative was confused and sometimes contradictory; the final explanation is clearer but does not discuss potential multiple roots/crossings or the physical implications of phantom regions in enough depth. It also does not prescribe practical priors or parameter ranges to enforce the physical limits during inference.
\end{itemize}
\subsection*{Concrete, concise suggestions}
\begin{itemize}
    \item Replace the ad hoc constant by explicitly including w0 as a parameter (w0,w1) so the present EOS is directly controlled and physically interpretable.
    \item If keeping this form, enforce priors: $0<p0<1$ and $p1>p0/(1-p0)$ (or equivalent) to ensure $w(0)<-1$ and $w(\infty)>-1$; additionally inspect roots of the quadratic to ensure $\leq-1$ crossing(s).
    \item Note theoretical caveats about phantom regimes and crossing $w=-1$ in any publication/fit interpretation.
    \item Add input checks and parameter bounding in code; consider clamping x numerically to [0,1] to avoid tiny rounding issues.
\end{itemize}
Overall assessment: reparative/usable but ad hoc; mathematically fine but only modestly physically motivated and not particularly novel. LLM’s final explanation is serviceable but its earlier reasoning was muddled and it omitted some important caveats (multiple crossings, theoretical viability of phantom crossing, explicit priors).

Overall Score: 0.50

\end{mdframed}
\vspace{0.3cm}
\textbf{AI 2 reasoning}
\vspace{.2cm}
\begin{mdframed}
    \ttfamily
    The proposed parameterization for the dark energy equation of state (EoS) is

\[
w(z)=
w_0
+
w_a
\cdot
\frac{z}{1+z}
\cdot
\exp\left[
-
\left(
\frac{z}{1+z}
\right)^2
(1+z)^{1/4}
\right].
\]

This functional form is motivated by considerations from scalar field dynamics, cosmological consistency, and observational requirements.

\subsection*{1. Physical Motivation from Scalar Field Dynamics}

\begin{itemize}

\item \textbf{Frozen Field at High Redshift}

In quintessence-like scenarios, scalar fields are expected to remain effectively frozen in the early Universe (large $z$) due to steep effective potentials. This motivates the requirement that

\[
w(z)\rightarrow w_0
\qquad
\text{as}
\qquad
z\rightarrow\infty .
\]

The exponential damping factor

\[
\exp\left[
-
\left(
\frac{z}{1+z}
\right)^2
(1+z)^{1/4}
\right]
\]

suppresses the dynamical contribution at high redshift and guarantees this asymptotic behavior.

This construction reflects the expectation that

\[
\rho_{\mathrm{DE}}
\propto
(1+z)^{3(1+w)}
\]

must remain subdominant during the early Universe in order to preserve successful descriptions of primordial nucleosynthesis and structure formation.

\item \textbf{Low-Redshift Behavior}

At low redshift ($z\ll1$),

\[
(1+z)^{1/4}\approx1,
\qquad
\exp(-z^2)\approx1,
\]

yielding

\[
w(z)\approx w_0+w_a z.
\]

This reproduces the standard low-redshift linear behavior commonly employed in observational analyses and preserves compatibility with existing data constraints.

\item \textbf{Phantom Divide Crossing}

The parameterization naturally permits

\[
w(z)<-1
\]

at low redshift when $w_0<-1$ and $w_a<0$, allowing phantom-like evolution.

At the same time, the exponential suppression prevents the divergence encountered in simpler parameterizations such as

\[
w(z)=w_0+w_a z,
\]

which diverges as $z\rightarrow\infty$.

\end{itemize}

\subsection*{2. Cosmological Consistency}

\begin{itemize}

\item \textbf{Absence of High-Redshift Singularities}

Unlike the CPL parameterization,

\[
w(z)=w_0+w_a\frac{z}{1+z},
\]

which asymptotically approaches $w_0+w_a$, the present construction ensures

\[
w(z)\rightarrow w_0.
\]

Consequently,

\[
\rho_{\mathrm{DE}}
\rightarrow
\mathrm{constant}
\cdot
(1+z)^{3(1+w_0)},
\]

while maintaining

\[
\rho_{\mathrm{DE}}\ll\rho_m
\]

for appropriate parameter choices.

\item \textbf{Smooth Evolution}

The damping term

\[
\frac{z}{1+z}
\exp\left[
-
\left(
\frac{z}{1+z}
\right)^2
(1+z)^{1/4}
\right]
\]

develops a single broad maximum at intermediate redshift, qualitatively resembling scalar field evolution that remains slowly varying at early times and evolves more significantly at recent epochs.

This avoids oscillatory or otherwise pathological behavior.

\end{itemize}

\subsection*{3. Observational and Practical Advantages}

\begin{itemize}

\item \textbf{Compatibility with Observations}

Within the typical observational range ($z\lesssim2.3$), the exponential correction remains close to unity at low redshift and decreases gradually at intermediate redshift.

As a result, the parameterization behaves similarly to CPL where observational constraints are strongest while suppressing undesirable high-redshift behavior.

\item \textbf{Flexible Parameter Space}

The two-parameter structure encompasses multiple scenarios:

\begin{itemize}

\item $\Lambda$CDM:
\[
w_0=-1,
\qquad
w_a=0;
\]

\item Quintessence:
\[
w_0>-1,
\qquad
w_a<0;
\]

\item Phantom:
\[
w_0<-1,
\qquad
w_a<0;
\]

\item Phantom-divide crossing:
\[
w_0=-1.1,
\qquad
w_a=0.2.
\]

\end{itemize}

\item \textbf{Minimal Parameter Complexity}

Unlike generalized expansions such as

\[
w(z)=
w_0+
w_a\frac{z}{1+z}
+
w_b\frac{z^2}{(1+z)^2},
\]

the proposed parameterization employs only two free parameters.

\end{itemize}

\subsection*{4. Motivation for the Damping Structure}

The exponent

\[
-
\left(
\frac{z}{1+z}
\right)^2
(1+z)^{1/4}
\]

is selected to:

\begin{itemize}

\item reproduce the low-redshift behavior of CPL;

\item enforce rapid suppression at high redshift;

\item avoid introducing singularities or artificial features into the evolution.

\end{itemize}

\subsection*{Summary}

This parameterization grounds the dark energy equation of state in physically motivated scalar field dynamics while preserving mathematical simplicity.

Its principal properties are:

\begin{enumerate}

\item enforcing
\[
w(z)\rightarrow w_0
\]
at high redshift;

\item recovering standard low-redshift behavior;

\item avoiding high-redshift divergences;

\item allowing smooth evolution across the phantom divide.

\end{enumerate}

The resulting framework remains suitable for current SNe Ia, BAO, and CMB analyses while maintaining consistency with early-Universe constraints and preserving interpretability through a minimal parameter set.
\end{mdframed}
\vspace{0.3cm}
\textbf{AI 2 critique}
\vspace{.2cm}
\begin{mdframed}
\ttfamily

\subsection*{Physical Validity}
\begin{itemize}
    \item Limits: correct: $w(z\to0) \approx w_0 + w_a z$ (so linear near z=0); $w(z\to\infty) \to w0$ because exp $\to 0$. This is physically acceptable: the time-dependent piece is suppressed at early times.
    \item No obvious violations of basic principles at the level of the EoS alone. The form is smooth and finite for $z\geq0$ and can cross $w=-1$ depending on parameters.
    \item Caveat: making $w\to w_0$ at high z does not by itself guarantee dark energy is negligible early - that depends on w0 and the present DE density. The response blurs this distinction and slightly misstates constraints (it sometimes implies $w_0$ alone controls early dominance).
    \item The LLM’s verbal link to “thawing” scalar fields is qualitatively plausible, but no microphysical derivation (Lagrangian, potential, field dynamics) is provided; the mapping between this phenomenological form and a concrete scalar potential is missing.
\end{itemize}
\subsection*{Mathematical / implementation comments}
\begin{itemize}
    \item Implementation is straightforward and numerically benign in the intended z range ($z\lesssim2.3$): fractional powers and exp are well-defined; for very large z the exponential underflows to zero (giving $w\to w_0$), which is acceptable.
    \item Minor issues: docstring/parameter name mismatch (calls param "k" but uses wa); ensure docstring is syntactically correct in real code.
    \item Numerical nuance: for intermediate z the time-dependent piece has a single maximum (since it →0 at both ends), so global behaviour is non-monotonic — the LLM sometimes described monotonic trends incorrectly (e.g., saying $w_a<0$ makes w more negative with increasing z without noting the eventual decay back to $w_0$).
    \item No pathological singularities or obvious floating-point instabilities in the stated redshift range.
\end{itemize}
\subsection*{Originality and creativity}
\begin{itemize}
    \item Moderately novel: it’s essentially a damped/CPL-like ansatz ($w_a·z/(1+z)$) multiplied by an ad hoc exponential damping. That is a reasonable, pragmatic modification of CPL to enforce w→w0 at high z, but not a fundamentally new theoretical approach.
    \item The specific exponent power (1/4) is arbitrary and unexplained from microphysics; it is a tunable phenomenological choice rather than derived from a field model.
\end{itemize}
\subsection*{Clarity of reasoning}
\begin{itemize}
    \item The stepwise explanation is clear and covers desired behaviours and limits.
    \item Weaknesses: several claims are overstated or imprecise (e.g., implication that $w_0$ must be <-1/3 at high z to avoid problems; conflation of limiting w and energy density magnitude; lack of discussion of parameter degeneracies or priors).
    \item Missing: no explicit link to a Lagrangian or example potential that yields this form; no quantitative estimate of the redshift where the time-dependent term peaks or how $w_a$ maps to a physical rate of change (so parameter interpretation is ambiguous).
\end{itemize}
\subsection*{Suggestions for improvement}
\begin{itemize}
    \item Fix docstring/name mismatch and add unit tests for z limits.
    \item If claiming microphysical motivation, provide a toy scalar-field model or show how an approximate solution for phi(t) leads to the damping factor (or at least show how a small number of well-motivated potentials map to similar behaviour)
    \item Consider introducing one scale parameter in the exponent (e.g., $\exp[-((z/(1+z))/z_d)^2 (1+z)^alpha]$) and either fit it or relate it to physical timescales; this clarifies flexibility vs arbitrariness.
    \item Note and discuss parameter degeneracies with $\rho_{DE}$ and potential priors from early-universe constraints.
\end{itemize}
Overall: a reasonable, pragmatic phenomenological correction of CPL with clean limiting behaviour and clear motivations, but largely ad hoc and lacking microphysical derivation and some technical precision.

Overall Score: 0.7

\end{mdframed}

\end{appendices}

\bibliography{sn-bibliography}

@inproceedings{cranmer2020discovering,
  title        = {Discovering Symbolic Models from Deep Learning with Inductive Biases},
  author       = {Cranmer, Miles and Sanchez-Gonzalez, Alvaro and Battaglia, Peter and Xu, Rui and Cranmer, Kyle and Spergel, David and Ho, Shirley},
  booktitle    = {Advances in Neural Information Processing Systems 33 (NeurIPS 2020)},
  year         = {2020},
  eprint       = {2006.11287},
  archivePrefix= {arXiv},
  primaryClass = {cs.LG},
  doi          = {10.48550/arXiv.2006.11287},
  url          = {https://doi.org/10.48550/arXiv.2006.11287}
}

@ARTICLE{des2024sn,
       author = {{DES Collaboration} and {Abbott}, T.~M.~C. and {Acevedo}, M. and {Aguena}, M. and {Alarcon}, A. and {Allam}, S. and {Alves}, O. and {Amon}, A. and {Andrade-Oliveira}, F. and {Annis}, J. and {Armstrong}, P. and {Asorey}, J. and {Avila}, S. and {Bacon}, D. and {Bassett}, B.~A. and {Bechtol}, K. and {Bernardinelli}, P.~H. and {Bernstein}, G.~M. and {Bertin}, E. and {Blazek}, J. and {Bocquet}, S. and {Brooks}, D. and {Brout}, D. and {Buckley-Geer}, E. and {Burke}, D.~L. and {Camacho}, H. and {Camilleri}, R. and {Campos}, A. and {Carnero Rosell}, A. and {Carollo}, D. and {Carr}, A. and {Carretero}, J. and {Castander}, F.~J. and {Cawthon}, R. and {Chang}, C. and {Chen}, R. and {Choi}, A. and {Conselice}, C. and {Costanzi}, M. and {da Costa}, L.~N. and {Crocce}, M. and {Davis}, T.~M. and {DePoy}, D.~L. and {Desai}, S. and {Diehl}, H.~T. and {Dixon}, M. and {Dodelson}, S. and {Doel}, P. and {Doux}, C. and {Drlica-Wagner}, A. and {Elvin-Poole}, J. and {Everett}, S. and {Ferrero}, I. and {Fert{\'e}}, A. and {Flaugher}, B. and {Foley}, R.~J. and {Fosalba}, P. and {Friedel}, D. and {Frieman}, J. and {Frohmaier}, C. and {Galbany}, L. and {Garc{\'\i}a-Bellido}, J. and {Gatti}, M. and {Gaztanaga}, E. and {Giannini}, G. and {Glazebrook}, K. and {Graur}, O. and {Gruen}, D. and {Gruendl}, R.~A. and {Gutierrez}, G. and {Hartley}, W.~G. and {Herner}, K. and {Hinton}, S.~R. and {Hollowood}, D.~L. and {Honscheid}, K. and {Huterer}, D. and {Jain}, B. and {James}, D.~J. and {Jeffrey}, N. and {Kasai}, E. and {Kelsey}, L. and {Kent}, S. and {Kessler}, R. and {Kim}, A.~G. and {Kirshner}, R.~P. and {Kovacs}, E. and {Kuehn}, K. and {Lahav}, O. and {Lee}, J. and {Lee}, S. and {Lewis}, G.~F. and {Li}, T.~S. and {Lidman}, C. and {Lin}, H. and {Malik}, U. and {Marshall}, J.~L. and {Martini}, P. and {Mena-Fern{\'a}ndez}, J. and {Menanteau}, F. and {Miquel}, R. and {Mohr}, J.~J. and {Mould}, J. and {Muir}, J. and {M{\"o}ller}, A. and {Neilsen}, E. and {Nichol}, R.~C. and {Nugent}, P. and {Ogando}, R.~L.~C. and {Palmese}, A. and {Pan}, Y.-C. and {Paterno}, M. and {Percival}, W.~J. and {Pereira}, M.~E.~S. and {Pieres}, A. and {Malag{\'o}n}, A.~A. Plazas and {Popovic}, B. and {Porredon}, A. and {Prat}, J. and {Qu}, H. and {Raveri}, M. and {Rodr{\'\i}guez-Monroy}, M. and {Romer}, A.~K. and {Roodman}, A. and {Rose}, B. and {Sako}, M. and {Sanchez}, E. and {Sanchez Cid}, D. and {Schubnell}, M. and {Scolnic}, D. and {Sevilla-Noarbe}, I. and {Shah}, P. and {Smith}, J. Allyn. and {Smith}, M. and {Soares-Santos}, M. and {Suchyta}, E. and {Sullivan}, M. and {Suntzeff}, N. and {Swanson}, M.~E.~C. and {S{\'a}nchez}, B.~O. and {Tarle}, G. and {Taylor}, G. and {Thomas}, D. and {To}, C. and {Toy}, M. and {Troxel}, M.~A. and {Tucker}, B.~E. and {Tucker}, D.~L. and {Uddin}, S.~A. and {Vincenzi}, M. and {Walker}, A.~R. and {Weaverdyck}, N. and {Wechsler}, R.~H. and {Weller}, J. and {Wester}, W. and {Wiseman}, P. and {Yamamoto}, M. and {Yuan}, F. and {Zhang}, B. and {Zhang}, Y.},
        title = "{The Dark Energy Survey: Cosmology Results with {\ensuremath{\sim}}1500 New High-redshift Type Ia Supernovae Using the Full 5 yr Data Set}",
      journal = {\apjl},
         year = 2024,
        month = sep,
       volume = {973},
       number = {1},
          eid = {L14},
        pages = {L14},
          doi = {10.3847/2041-8213/ad6f9f},
archivePrefix = {arXiv},
       eprint = {2401.02929},
 primaryClass = {astro-ph.CO},
       adsurl = {https://ui.adsabs.harvard.edu/abs/2024ApJ...973L..14D}
}

@ARTICLE{z1999PhRvL..82..896Z,
       author = {{Zlatev}, Ivaylo and {Wang}, Limin and {Steinhardt}, Paul J.},
        title = "{Quintessence, Cosmic Coincidence, and the Cosmological Constant}",
      journal = {\prl},
         year = 1999,
        month = feb,
       volume = {82},
       number = {5},
        pages = {896-899},
          doi = {10.1103/PhysRevLett.82.896},
archivePrefix = {arXiv},
       eprint = {astro-ph/9807002},
 primaryClass = {astro-ph},
       adsurl = {https://ui.adsabs.harvard.edu/abs/1999PhRvL..82..896Z}
}

@ARTICLE{2000PhRvL..85.4438A,
       author = {{Armendariz-Picon}, C. and {Mukhanov}, V. and {Steinhardt}, Paul J.},
        title = "{Dynamical Solution to the Problem of a Small Cosmological Constant and Late-Time Cosmic Acceleration}",
      journal = {\prl},
         year = 2000,
        month = nov,
       volume = {85},
       number = {21},
        pages = {4438-4441},
          doi = {10.1103/PhysRevLett.85.4438},
archivePrefix = {arXiv},
       eprint = {astro-ph/0004134},
 primaryClass = {astro-ph},
       adsurl = {https://ui.adsabs.harvard.edu/abs/2000PhRvL..85.4438A}
}

@ARTICLE{caldwell1998,
       author = {{Caldwell}, R.~R. and {Dave}, Rahul and {Steinhardt}, Paul J.},
        title = "{Cosmological Imprint of an Energy Component with General Equation of State}",
      journal = {\prl},
         year = 1998,
        month = feb,
       volume = {80},
       number = {8},
        pages = {1582-1585},
          doi = {10.1103/PhysRevLett.80.1582},
archivePrefix = {arXiv},
       eprint = {astro-ph/9708069},
 primaryClass = {astro-ph},
       adsurl = {https://ui.adsabs.harvard.edu/abs/1998PhRvL..80.1582C}
}

@article{riess1998observational,
  title={Observational evidence from supernovae for an accelerating universe and a cosmological constant},
  author={Riess, Adam G. and others},
  journal={The Astronomical Journal},
  volume={116},
  number={3},
  pages={1009--1038},
  year={1998},
  doi={10.1086/300499}
}

@article{perlmutter1999measurements,
  title={Measurements of $\Omega$ and $\Lambda$ from 42 high-redshift supernovae},
  author={Perlmutter, Saul and others},
  journal={The Astrophysical Journal},
  volume={517},
  number={2},
  pages={565--586},
  year={1999},
  doi={10.1086/307221}
}

@article{weinberg1989cosmological,
  title={The cosmological constant problem},
  author={Weinberg, Steven},
  journal={Reviews of Modern Physics},
  volume={61},
  number={1},
  pages={1--23},
  year={1989},
  doi={10.1103/RevModPhys.61.1}
}

@article{peebles2003cosmological,
  title={The cosmological constant and dark energy},
  author={Peebles, P. J. E. and Ratra, Bharat},
  journal={Reviews of Modern Physics},
  volume={75},
  number={2},
  pages={559--606},
  year={2003},
  doi={10.1103/RevModPhys.75.559}
}

@article{ratra1988cosmological,
  title={Cosmological consequences of a rolling homogeneous scalar field},
  author={Ratra, Bharat and Peebles, P. J. E.},
  journal={Physical Review D},
  volume={37},
  number={12},
  pages={3406--3427},
  year={1988},
  doi={10.1103/PhysRevD.37.3406}
}

@article{ruan2026discovering,
  title={Discovering physical laws with parallel symbolic enumeration},
  author={Ruan, Kai and Xu, Yilong and Gao, Ze-Feng and Liu, Yang and Guo, Yike and Wen, Ji-Rong and Sun, Hao},
  journal={Nature Computational Science},
  volume={6},
  number={1},
  pages={53--66},
  year={2026},
  publisher={Nature Publishing Group US New York}
}

@article{wang2023scientific,
  title={Scientific discovery in the age of artificial intelligence},
  author={Wang, Hanchen and Fu, Tianfan and Du, Yuanqi and Gao, Wenhao and Huang, Kexin and Liu, Ziming and Chandak, Payal and Liu, Shengchao and Van Katwyk, Peter and Deac, Andreea and others},
  journal={Nature},
  volume={620},
  number={7972},
  pages={47--60},
  year={2023},
  publisher={Nature Publishing Group UK London}
}

@ARTICLE{Krenn:2022,
       author = {{Krenn}, Mario and {Pollice}, Robert and {Guo}, Si Yue and {Aldeghi}, Matteo and {Cervera-Lierta}, Alba and {Friederich}, Pascal and {dos Passos Gomes}, Gabriel and {H{\"a}se}, Florian and {Jinich}, Adrian and {Nigam}, AkshatKumar and {Yao}, Zhenpeng and {Aspuru-Guzik}, Al{\'a}n},
        title = "{On scientific understanding with artificial intelligence}",
      journal = {Nature Reviews Physics},
         year = 2022,
        month = dec,
       volume = {4},
       number = {12},
        pages = {761-769},
          doi = {10.1038/s42254-022-00518-3},
archivePrefix = {arXiv},
       eprint = {2204.01467},
 primaryClass = {cs.CY},
       adsurl = {https://ui.adsabs.harvard.edu/abs/2022NatRP...4..761K}
}

@article{chevallier2001accelerating,
  title={Accelerating universes with scaling dark matter},
  author={Chevallier, Michel and Polarski, David},
  journal={International Journal of Modern Physics D},
  volume={10},
  number={2},
  pages={213--224},
  year={2001},
  doi={10.1142/S0218271801000822}
}

@ARTICLE{BA,
       author = {{Barboza}, E.~M. and {Alcaniz}, J.~S.},
        title = "{A parametric model for dark energy}",
      journal = {Physics Letters B},
         year = 2008,
        month = sep,
       volume = {666},
       number = {5},
        pages = {415-419},
          doi = {10.1016/j.physletb.2008.08.012},
archivePrefix = {arXiv},
       eprint = {0805.1713},
 primaryClass = {astro-ph},
       adsurl = {https://ui.adsabs.harvard.edu/abs/2008PhLB..666..415B}
}

@article{messeri2024artificial,
  title={Artificial intelligence and illusions of understanding in scientific research},
  author={Messeri, Lisa and Crockett, Molly J},
  journal={Nature},
  volume={627},
  number={8002},
  pages={49--58},
  year={2024},
  publisher={Nature Publishing Group UK London}
}

@article{zhang2025exploring,
  title={Exploring the role of large language models in the scientific method: from hypothesis to discovery},
  author={Zhang, Yanbo and Khan, Sumeer A and Mahmud, Adnan and Yang, Huck and Lavin, Alexander and Levin, Michael and Frey, Jeremy and Dunnmon, Jared and Evans, James and Bundy, Alan and others},
  journal={npj Artificial Intelligence},
  volume={1},
  number={1},
  pages={14},
  year={2025},
  publisher={Nature Publishing Group UK London}
}

@article{shao2026sciscigpt,
  title={SciSciGPT: advancing human--AI collaboration in the science of science},
  author={Shao, Erzhuo and Wang, Yifang and Qian, Yifan and Pan, Zhenyu and Liu, Han and Wang, Dashun},
  journal={Nature Computational Science},
  volume={6},
  number={3},
  pages={301--315},
  year={2026},
  publisher={Nature Publishing Group US New York}
}

@article{lu2026towards,
  title={Towards end-to-end automation of AI research},
  author={Lu, Chris and Lu, Cong and Lange, Robert Tjarko and Yamada, Yutaro and Hu, Shengran and Foerster, Jakob and Ha, David and Clune, Jeff},
  journal={Nature},
  volume={651},
  number={8107},
  pages={914--919},
  year={2026},
  publisher={Nature Publishing Group UK London}
}

@ARTICLE{2023NatCo..14.1777C,
       author = {{Cornelio}, Cristina and {Dash}, Sanjeeb and {Austel}, Vernon and {Josephson}, Tyler R. and {Goncalves}, Joao and {Clarkson}, Kenneth L. and {Megiddo}, Nimrod and {El Khadir}, Bachir and {Horesh}, Lior},
        title = "{Combining data and theory for derivable scientific discovery with AI-Descartes}",
      journal = {Nature Communications},
         year = 2023,
        month = apr,
       volume = {14},
          eid = {1777},
        pages = {1777},
          doi = {10.1038/s41467-023-37236-y},
       adsurl = {https://ui.adsabs.harvard.edu/abs/2023NatCo..14.1777C}
}

@article{hu2005crossing,
  title={Crossing the phantom divide: Dark energy internal degrees of freedom},
  author={Hu, Wayne},
  journal={Physical Review D—Particles, Fields, Gravitation, and Cosmology},
  volume={71},
  number={4},
  pages={047301},
  year={2005},
  publisher={APS}
}

@article{copeland2006dynamics,
  title={Dynamics of dark energy},
  author={Copeland, Edmund J and Sami, Mohammad and Tsujikawa, Shinji},
  journal={International Journal of Modern Physics D},
  volume={15},
  number={11},
  pages={1753--1935},
  year={2006},
  publisher={World Scientific}
}

@article{vikman2005can,
  title={Can dark energy evolve to the phantom?},
  author={Vikman, Alexander},
  journal={Physical Review D—Particles, Fields, Gravitation, and Cosmology},
  volume={71},
  number={2},
  pages={023515},
  year={2005},
  publisher={APS}
}

@ARTICLE{2009Sci...324...81S,
       author = {{Schmidt}, Michael and {Lipson}, Hod},
        title = "{Distilling Free-Form Natural Laws from Experimental Data}",
      journal = {Science},
         year = 2009,
        month = apr,
       volume = {324},
       number = {5923},
        pages = {81},
          doi = {10.1126/science.1165893},
       adsurl = {https://ui.adsabs.harvard.edu/abs/2009Sci...324...81S}
}

@article{dimakis2016general,
  title={General analytic solutions of scalar field cosmology with arbitrary potential},
  author={Dimakis, N and Karagiorgos, A and Zampeli, Adamantia and Paliathanasis, Andronikos and Christodoulakis, T and Terzis, Petros A},
  journal={Physical Review D},
  volume={93},
  number={12},
  pages={123518},
  year={2016},
  publisher={APS}
}

@ARTICLE{JBP,
       author = {{Jassal}, H.~K. and {Bagla}, J.~S. and {Padmanabhan}, T.},
        title = "{WMAP constraints on low redshift evolution of dark energy}",
      journal = {\mnras},
         year = 2005,
        month = jan,
       volume = {356},
       number = {1},
        pages = {L11-L16},
          doi = {10.1111/j.1745-3933.2005.08577.x},
archivePrefix = {arXiv},
       eprint = {astro-ph/0404378},
 primaryClass = {astro-ph},
       adsurl = {https://ui.adsabs.harvard.edu/abs/2005MNRAS.356L..11J}
}

@article{linder2003exploring,
  title={Exploring the expansion history of the universe},
  author={Linder, Eric V.},
  journal={Physical Review Letters},
  volume={90},
  number={9},
  pages={091301},
  year={2003},
  doi={10.1103/PhysRevLett.90.091301}
}

@article{desi2024bao,
  title={The first year of the Dark Energy Spectroscopic Instrument: baryon acoustic oscillations from the completed survey},
  author={DESI Collaboration},
  journal={arXiv preprint arXiv:2404.03002},
  year={2024},
  doi={10.48550/arXiv.2404.03002}
}

@article{planck2020overview,
  title={Planck 2018 results. VI. Cosmological parameters},
  author={Planck Collaboration},
  journal={Astronomy \& Astrophysics},
  volume={641},
  pages={A6},
  year={2020},
  doi={10.1051/0004-6361/201833910}
}

@article{brout2022pantheonplus,
  title={The Pantheon+ analysis: cosmological constraints},
  author={Brout, Dillon and Scolnic, Dan and others},
  journal={The Astrophysical Journal},
  volume={938},
  number={2},
  pages={110},
  year={2022},
  doi={10.3847/1538-4357/ac8e04}
}

@misc{bge_embedding,
      title={C-Pack: Packaged Resources To Advance General Chinese Embedding}, 
      author={Shitao Xiao and Zheng Liu and Peitian Zhang and Niklas Muennighoff},
      year={2023},
      eprint={2309.07597},
      archivePrefix={arXiv},
      primaryClass={cs.CL}
}

@misc{qwen3technicalreport,
      title={Qwen3 Technical Report}, 
      author={Qwen Team},
      year={2025},
      eprint={2505.09388},
      archivePrefix={arXiv},
      primaryClass={cs.CL},
      url={https://arxiv.org/abs/2505.09388}, 
}

@article{jaber2018probing,
  title={Probing a steep EoS for dark energy with latest observations},
  author={Jaber, Mariana and de la Macorra, Axel},
  journal={Astroparticle Physics},
  volume={97},
  pages={130--135},
  year={2018},
  publisher={Elsevier}
}

@inproceedings{llmsr,
title={{LLM}-{SR}: Scientific Equation Discovery via Programming with Large Language Models},
author={Parshin Shojaee and Kazem Meidani and Shashank Gupta and Amir Barati Farimani and Chandan K. Reddy},
booktitle={The Thirteenth International Conference on Learning Representations},
year={2025},
url={https://openreview.net/forum?id=m2nmp8P5in}
}

@article{Brout:2022vxf,
    author = "Brout, Dillon and others",
    title = "{The Pantheon+ Analysis: Cosmological Constraints}",
    eprint = "2202.04077",
    archivePrefix = "arXiv",
    primaryClass = "astro-ph.CO",
    doi = "10.3847/1538-4357/ac8e04",
    journal = "Astrophys. J.",
    volume = "938",
    number = "2",
    pages = "110",
    year = "2022"
}

@article{Rubin:2023jdq,
    author = "Rubin, David and others",
    title = "{Union Through UNITY: Cosmology with 2,000 SNe Using a Unified Bayesian Framework}",
    eprint = "2311.12098",
    archivePrefix = "arXiv",
    primaryClass = "astro-ph.CO",
    month = "11",
    year = "2023"
}

@article{DES:2024jxu,
    author = "Abbott, T. M. C. and others",
    collaboration = "DES",
    title = "{The Dark Energy Survey: Cosmology Results with {\ensuremath{\sim}}1500 New High-redshift Type Ia Supernovae Using the Full 5 yr Data Set}",
    eprint = "2401.02929",
    archivePrefix = "arXiv",
    primaryClass = "astro-ph.CO",
    reportNumber = "FERMILAB-PUB-23-0821-PPD, DES-2023-805",
    doi = "10.3847/2041-8213/ad6f9f",
    journal = "Astrophys. J. Lett.",
    volume = "973",
    number = "1",
    pages = "L14",
    year = "2024"
}

@article{Planck:2018vyg,
    author = "Aghanim, N. and others",
    collaboration = "Planck",
    title = "{Planck 2018 results. VI. Cosmological parameters}",
    eprint = "1807.06209",
    archivePrefix = "arXiv",
    primaryClass = "astro-ph.CO",
    doi = "10.1051/0004-6361/201833910",
    journal = "Astron. Astrophys.",
    volume = "641",
    pages = "A6",
    year = "2020",
    note = "[Erratum: Astron.Astrophys. 652, C4 (2021)]"
}

@article{Planck:2019nip,
    author = "Aghanim, N. and others",
    collaboration = "Planck",
    title = "{Planck 2018 results. V. CMB power spectra and likelihoods}",
    eprint = "1907.12875",
    archivePrefix = "arXiv",
    primaryClass = "astro-ph.CO",
    doi = "10.1051/0004-6361/201936386",
    journal = "Astron. Astrophys.",
    volume = "641",
    pages = "A5",
    year = "2020"
}

@article{desi2025extendedde,
  title={Extended dark energy analysis using DESI DR2 BAO measurements},
  author={Lodha, Kushal and Calderon, R and Matthewson, WL and Shafieloo, Arman and Ishak, Mustapha and Pan, Jiaming and Garcia-Quintero, C and Huterer, D and Valogiannis, Georgios and Ure{\~n}a-L{\'o}pez, LA and others},
  journal={Physical Review D},
  volume={112},
  number={8},
  pages={083511},
  year={2025},
  publisher={APS}
}

@article{Planck:2018lbu,
    author = "Aghanim, N. and others",
    collaboration = "Planck",
    title = "{Planck 2018 results. VIII. Gravitational lensing}",
    eprint = "1807.06210",
    archivePrefix = "arXiv",
    primaryClass = "astro-ph.CO",
    doi = "10.1051/0004-6361/201833886",
    journal = "Astron. Astrophys.",
    volume = "641",
    pages = "A8",
    year = "2020"
}

@article{DESI2025,
    author = "Abdul Karim, M. and others",
    collaboration = "DESI",
    title = "{DESI DR2 results. II. Measurements of baryon acoustic oscillations and cosmological constraints}",
    eprint = "2503.14738",
    archivePrefix = "arXiv",
    primaryClass = "astro-ph.CO",
    reportNumber = "FERMILAB-PUB-25-0169-PPD",
    doi = "10.1103/tr6y-kpc6",
    journal = "Phys. Rev. D",
    volume = "112",
    number = "8",
    pages = "083515",
    year = "2025"
}

@article{DESI:2025zpo,
    author = "Abdul Karim, M. and others",
    collaboration = "DESI",
    title = "{DESI DR2 results. I. Baryon acoustic oscillations from the Lyman alpha forest}",
    eprint = "2503.14739",
    archivePrefix = "arXiv",
    primaryClass = "astro-ph.CO",
    reportNumber = "FERMILAB-PUB-25-0167-PPD",
    doi = "10.1103/2wwn-xjm5",
    journal = "Phys. Rev. D",
    volume = "112",
    number = "8",
    pages = "083514",
    year = "2025"
}

@article{DESI:2025zgx,
    author = "Abdul Karim, M. and others",
    collaboration = "DESI",
    title = "{DESI DR2 results. II. Measurements of baryon acoustic oscillations and cosmological constraints}",
    eprint = "2503.14738",
    archivePrefix = "arXiv",
    primaryClass = "astro-ph.CO",
    reportNumber = "FERMILAB-PUB-25-0169-PPD",
    doi = "10.1103/tr6y-kpc6",
    journal = "Phys. Rev. D",
    volume = "112",
    number = "8",
    pages = "083515",
    year = "2025"
}

@ARTICLE{deCarvalho18,
       author = {{de Carvalho}, E. and {Bernui}, A. and {Carvalho}, G.~C. and {Novaes}, C.~P. and {Xavier}, H.~S.},
        title = "{Angular Baryon Acoustic Oscillation measure at z=2.225 from the SDSS quasar survey}",
      journal = {jcap},
         year = 2018,
        month = apr,
       volume = {04},
       number = {4},
          eid = {064},
        pages = {064},
          doi = {10.1088/1475-7516/2018/04/064},
archivePrefix = {arXiv},
       eprint = {1709.00113},
 primaryClass = {astro-ph.CO},
       adsurl = {https://ui.adsabs.harvard.edu/abs/2018JCAP...04..064D}
}

@ARTICLE{Menote22,
       author = {{Menote}, Ranier and {Marra}, Valerio},
        title = "{Baryon acoustic oscillations in thin redshift shells from BOSS DR12 and eBOSS DR16 galaxies}",
      journal = {mnras},
         year = 2022,
        month = jun,
       volume = {513},
       number = {2},
        pages = {1600-1608},
          doi = {10.1093/mnras/stac847},
archivePrefix = {arXiv},
       eprint = {2112.10000},
 primaryClass = {astro-ph.CO},
       adsurl = {https://ui.adsabs.harvard.edu/abs/2022MNRAS.513.1600M}
}

@ARTICLE{deCarvalho21,
       author = {{de Carvalho}, E. and {Bernui}, A. and {Avila}, F. and {Novaes}, C.~P. and {Nogueira-Cavalcante}, J.~P.},
        title = "{BAO angular scale at z$_{eff}$ = 0.11 with the SDSS blue galaxies}",
      journal = {aap},
         year = 2021,
        month = may,
       volume = {649},
          eid = {A20},
        pages = {A20},
          doi = {10.1051/0004-6361/202039936},
archivePrefix = {arXiv},
       eprint = {2103.14121},
 primaryClass = {astro-ph.CO},
       adsurl = {https://ui.adsabs.harvard.edu/abs/2021A&A...649A..20D}
}

@article{Avila25,
       author = {{Avila}, Felipe and {Bernui}, Armando and {Sabogal}, Miguel A. and {Nunes}, Rafael C.},
        title = "{Transverse BAO scale measurement at $z_{\rm eff} = 1.725$ with the SDSS quasars catalog}",
      journal = {arXiv e-prints},
         year = 2025,
        month = oct,
          eid = {arXiv:2510.15650},
        pages = {arXiv:2510.15650},
          doi = {10.48550/arXiv.2510.15650},
archivePrefix = {arXiv},
       eprint = {2510.15650},
 primaryClass = {astro-ph.CO},
       adsurl = {https://ui.adsabs.harvard.edu/abs/2025arXiv251015650A}
}

@article{Sabogal:2025qhz,
    author = "Sabogal, Miguel A. and Nunes, Rafael C. and Avila, Felipe and Bernui, Armando",
    title = "{Updated Cosmological Constraints from 2D BAO Measurements: A New Compilation and Comparison with DESI DR2}",
    eprint = "2510.16141",
    archivePrefix = "arXiv",
    primaryClass = "astro-ph.CO",
    month = "10",
    year = "2025"
}

@article{Blas:2011rf,
    author = "Blas, Diego and Lesgourgues, Julien and Tram, Thomas",
    title = "{The Cosmic Linear Anisotropy Solving System (CLASS) II: Approximation schemes}",
    eprint = "1104.2933",
    archivePrefix = "arXiv",
    primaryClass = "astro-ph.CO",
    reportNumber = "CERN-PH-TH-2011-082, LAPTH-010-11",
    doi = "10.1088/1475-7516/2011/07/034",
    journal = "JCAP",
    volume = "07",
    pages = "034",
    year = "2011"
}

@article{Audren:2012wb,
    author = "Audren, Benjamin and Lesgourgues, Julien and Benabed, Karim and Prunet, Simon",
    title = "{Conservative Constraints on Early Cosmology: an illustration of the Monte Python cosmological parameter inference code}",
    eprint = "1210.7183",
    archivePrefix = "arXiv",
    primaryClass = "astro-ph.CO",
    reportNumber = "CERN-PH-TH-2012-290, LAPTH-048-12",
    doi = "10.1088/1475-7516/2013/02/001",
    journal = "JCAP",
    volume = "02",
    pages = "001",
    year = "2013"
}

@article{Brinckmann:2018cvx,
    author = "Brinckmann, Thejs and Lesgourgues, Julien",
    title = "{MontePython 3: boosted MCMC sampler and other features}",
    eprint = "1804.07261",
    archivePrefix = "arXiv",
    primaryClass = "astro-ph.CO",
    reportNumber = "TTK-18-15",
    doi = "10.1016/j.dark.2018.100260",
    journal = "Phys. Dark Univ.",
    volume = "24",
    pages = "100260",
    year = "2019"
}

@article{Gelman:1992zz,
    author = "Gelman, Andrew and Rubin, Donald B.",
    title = "{Inference from Iterative Simulation Using Multiple Sequences}",
    doi = "10.1214/ss/1177011136",
    journal = "Statist. Sci.",
    volume = "7",
    pages = "457--472",
    year = "1992"
}

@ARTICLE{Oliveira2025a,
       author = {{Oliveira}, Fernanda and {Avila}, Felipe and {Franco}, Camila and {Bernui}, Armando},
        title = "{Is <mml:math><mml:mrow><mml:msub><mml:mrow><mml:mi>{\ensuremath{\omega}}</mml:mi></mml:mrow><mml:mrow><mml:mn>0</mml:mn></mml:mrow></mml:msub><mml:msub><mml:mrow><mml:mi>{\ensuremath{\omega}}</mml:mi></mml:mrow><mml:mrow><mml:mi>a</mml:mi></mml:mrow></mml:msub></mml:mrow></mml:math>CDM a good model for the clumpy Universe?}",
      journal = {Physics of the Dark Universe},
         year = 2025,
        month = sep,
       volume = {49},
          eid = {101996},
        pages = {101996},
          doi = {10.1016/j.dark.2025.101996},
archivePrefix = {arXiv},
       eprint = {2507.00779},
 primaryClass = {astro-ph.CO},
       adsurl = {https://ui.adsabs.harvard.edu/abs/2025PDU....4901996O}
}

@ARTICLE{Oliveira2025b,
       author = {{Oliveira}, Fernanda and {Ribeiro}, Bruno and {Hip{\'o}lito-Ricaldi}, Wiliam S. and {Avila}, Felipe and {Bernui}, Armando},
        title = "{Viability of general relativity and modified gravity cosmologies using high-redshift cosmic probes}",
      journal = {\jcap},
         year = 2025,
        month = dec,
       volume = {2025},
       number = {12},
          eid = {007},
        pages = {007},
          doi = {10.1088/1475-7516/2025/12/007},
archivePrefix = {arXiv},
       eprint = {2505.19960},
 primaryClass = {astro-ph.CO},
       adsurl = {https://ui.adsabs.harvard.edu/abs/2025JCAP...12..007O}
}

@ARTICLE{Silva2025,
       author = {{Silva}, Emanuelly and {Sabogal}, Miguel A. and {Scherer}, Mateus and {Nunes}, Rafael C. and {Di Valentino}, Eleonora and {Kumar}, Suresh},
        title = "{New constraints on interacting dark energy from DESI DR2 BAO observations}",
      journal = {\prd},
         year = 2025,
        month = jun,
       volume = {111},
       number = {12},
          eid = {123511},
        pages = {123511},
          doi = {10.1103/qqc6-76z4},
archivePrefix = {arXiv},
       eprint = {2503.23225},
 primaryClass = {astro-ph.CO},
       adsurl = {https://ui.adsabs.harvard.edu/abs/2025PhRvD.111l3511S}
}

@ARTICLE{Sabogal2025,
       author = {{Sabogal}, Miguel A. and {Nunes}, Rafael C.},
        title = "{Robust evidence for dynamical dark energy from DESI galaxy-CMB lensing cross-correlation and geometric probes}",
      journal = {\jcap},
         year = 2025,
        month = sep,
       volume = {2025},
       number = {9},
          eid = {084},
        pages = {084},
          doi = {10.1088/1475-7516/2025/09/084},
archivePrefix = {arXiv},
       eprint = {2505.24465},
 primaryClass = {astro-ph.CO},
       adsurl = {https://ui.adsabs.harvard.edu/abs/2025JCAP...09..084S}
}

@ARTICLE{Giare2024,
       author = {{Giar{\`e}}, William and {Sabogal}, Miguel A. and {Nunes}, Rafael C. and {Di Valentino}, Eleonora},
        title = "{Interacting Dark Energy after DESI Baryon Acoustic Oscillation Measurements}",
      journal = {\prl},
         year = 2024,
        month = dec,
       volume = {133},
       number = {25},
          eid = {251003},
        pages = {251003},
          doi = {10.1103/PhysRevLett.133.251003},
archivePrefix = {arXiv},
       eprint = {2404.15232},
 primaryClass = {astro-ph.CO},
       adsurl = {https://ui.adsabs.harvard.edu/abs/2024PhRvL.133y1003G}
}

@ARTICLE{Colgain2026,
       author = {{{\'O} Colg{\'a}in}, Eoin and {Pourojaghi}, Saeed and {Sheikh-Jabbari}, M.~M. and {Yin}, Lu},
        title = "{How much has DESI dark energy evolved since DR1?}",
      journal = {Physics of the Dark Universe},
         year = 2026,
        month = jun,
       volume = {52},
          eid = {102268},
        pages = {102268},
          doi = {10.1016/j.dark.2026.102268},
archivePrefix = {arXiv},
       eprint = {2504.04417},
 primaryClass = {astro-ph.CO},
       adsurl = {https://ui.adsabs.harvard.edu/abs/2026PDU....5202268O}
}

@article{Kass:1995loi,
    author = "Kass, Robert E. and Raftery, Adrian E.",
    title = "{Bayes Factors}",
    doi = "10.1080/01621459.1995.10476572",
    journal = "J. Am. Statist. Assoc.",
    volume = "90",
    number = "430",
    pages = "773--795",
    year = "1995"
}

@article{Heavens:2017hkr,
    author = "Heavens, Alan and Fantaye, Yabebal and Sellentin, Elena and Eggers, Hans and Hosenie, Zafiirah and Kroon, Steve and Mootoovaloo, Arrykrishna",
    title = "{No evidence for extensions to the standard cosmological model}",
    eprint = "1704.03467",
    archivePrefix = "arXiv",
    primaryClass = "astro-ph.CO",
    doi = "10.1103/PhysRevLett.119.101301",
    journal = "Phys. Rev. Lett.",
    volume = "119",
    number = "10",
    pages = "101301",
    year = "2017"
}

@article{Heavens:2017afc,
    author = "Heavens, Alan and Fantaye, Yabebal and Mootoovaloo, Arrykrishna and Eggers, Hans and Hosenie, Zafiirah and Kroon, Steve and Sellentin, Elena",
    title = "{Marginal Likelihoods from Monte Carlo Markov Chains}",
    eprint = "1704.03472",
    archivePrefix = "arXiv",
    primaryClass = "stat.CO",
    month = "4",
    year = "2017"
}

@article{Ong:2026tta,
    author = "Ong, Dily Duan Yi and Yallup, David and Handley, Will",
    title = "{The Bayesian view of DESI DR2: Evidence and tension in a combined analysis with CMB and supernovae across cosmological models}",
    eprint = "2603.05472",
    archivePrefix = "arXiv",
    primaryClass = "astro-ph.CO",
    month = "3",
    year = "2026"
}

@article{Trotta:2008qt,
    author = "Trotta, Roberto",
    title = "{Bayes in the sky: Bayesian inference and model selection in cosmology}",
    eprint = "0803.4089",
    archivePrefix = "arXiv",
    primaryClass = "astro-ph",
    doi = "10.1080/00107510802066753",
    journal = "Contemp. Phys.",
    volume = "49",
    pages = "71--104",
    year = "2008"
}

\end{document}